\documentclass[showpacs,preprintnumbers,amsmath,amssymb, prd, nofootinbib]{revtex4}
\pdfoutput=1
\usepackage{amsmath,amssymb,graphicx}
\usepackage{epsf}
\usepackage{epstopdf}
\usepackage{color}
\usepackage {amssymb}
\newcommand{\nc}{\newcommand}
\nc{\ba}{\begin{eqnarray}}
\nc{\ea}{\end{eqnarray}}
\newcommand\be{\begin{equation}}
\newcommand\ee{\end{equation}}

\nc{\e}{{\bf{e}}}
\nc{\kk}{{\bf{k}}}
\nc{\pp}{{\bf{p}}}
\nc{\cM}{{\cal{M}}}

\begin{document}

%%%%%%%%%%%%%%%%%%%%%%%%%%%%%%%%%%%%%%%%%%%%%%%%%%%%%%%%%%%%%%%%%%%%%%%
\title{Curvature Perturbations in Anisotropic Inflation with Symmetry Breaking}

\author{Razieh Emami}
\email{emami@ipm.ir}
\affiliation{School of Physics, Institute for Research in Fundamental Sciences (IPM), P. O. Box 19395-5531, Tehran, Iran}

\author{Hassan Firouzjahi}
\email{firouz@ipm.ir}
\affiliation{School of Astronomy, Institute for Research in Fundamental Sciences (IPM), P. O. Box 19395-5531, Tehran, Iran}

\preprint{  IPM/A-2012/017 }

\begin{abstract}
We study  curvature perturbations in the anisotropic inflationary model  with a complex scalar field charged under a $U(1)$ gauge field in Bianchi I universe. Due to Abelian Higgs mechanism, the gauge field receives an additional longitudinal mode. We verify that the dominant contributions into statistical  anisotropies come from matter fields perturbations and one can neglect the contributions from the metric perturbations. It is shown that the contribution of longitudinal mode into the statistical anisotropy power spectrum, though exponentially small,
has an opposite sign compared to the corresponding contribution from the transverse mode.
We obtain an upper bound on gauge coupling in order to satisfy the observational constraints
on curvature perturbations anisotropy.

%Keywords :  Gravity Waves,
\end{abstract}

\maketitle

%%%%%%%%%%%%%%%%%%%%%%%%%%%%%%%%%%%%%%%%%%%%%%%%%%
\section{Introduction}

Simple models of inflation predict almost scale invariant, almost adiabatic and almost Gaussian perturbations on Cosmic Microwave Background (CMB) which are in very good agreements with cosmological observations \cite{Komatsu:2010fb}. There may be indications of statistical anisotropies on CMB \cite{Hanson:2009gu, Hanson:2010gu} which can not be generated in simple models inflation based on scalar fields. Although the statistical significance of the possible statistical anisotropies on CMB is not high, nonetheless this opens up the interesting possibilities that  primordial seeds in generating curvature perturbations during inflation may not be statistically isotropic. This will  shed new light on the mechanisms of inflation.

One can parameterize the statistical anisotropy via \cite{Ackerman:2007nb}
$P_{\zeta} (\vec k) = P_0(k) \left(  1+ g_* \cos^2 \theta  \right)$ in which $P_{\zeta} (\vec k) $
represents the curvature perturbations and $\theta$ is the angle between the preferred direction in the sky which breaks the rotational invariance and the momentum vector $\vec k$.  Constraints from CMB and large scale structure indicate that $| g_*| \lesssim 0.4$
\cite{Groeneboom:2009cb, Pullen:2010zy}.

Motivated by these observations, there have been many attempts in the literature to generate primordial anisotropies during inflation. The natural way to break the statistical anisotropy during inflation is to employ a gauge field or a vector field to seed the anisotropies at the order of few percent which may be detectable on CMB \cite{Ford:1989me, Kaloper:1991rw, Kawai:1998bn,  Barrow:2005qv, Barrow:2009gx, Campanelli:2009tk, Golovnev:2008cf, Kanno:2008gn, Pitrou:2008gk, Moniz:2010cm, Boehmer:2007ut, Koivisto:2008xf, Maleknejad:2011jr, Maleknejad:2012as, Yokoyama:2008xw, Emami:2011yi,   Lyth:2012vn,   Dimopoulos:2009vu, Dimopoulos:2012av,
Dimastrogiovanni:2010sm, ValenzuelaToledo:2009af}.

As demonstrated in \cite{Himmetoglu:2008zp},  in models employing a massive vector field in which the gauge invariance is broken explicitly, the excitations contain a ghost   which is not acceptable physically. Therefore, it is crucial that the vector field is protected by a gauge symmetry so the longitudinal mode of the vector field excitations is not physical. On the other hand, because of the conformal invariance of models with gauge fields, any excitation of gauge field during inflation is diluted and can not seed the desired anisotropies. Therefor it is essential that one breaks the conformal invariance while keeping the gauge symmetry explicit. This approach was employed in different contexts in
\cite{Turner:1987bw, Ratra:1991bn, Demozzi:2009fu, Martin:2007ue, Emami:2009vd, Kanno:2009ei, Caldwell:2011ra, Jain:2012ga, Jain:2012vm,
Watanabe:2009ct, Emami:2010rm, Kanno:2010nr, Murata:2011wv, Bhowmick:2011em, Hervik:2011xm, Thorsrud:2012mu, Dimopoulos:2010xq, Yamamoto:2012tq}. Specifically, one can consider the model in which the $U(1)$ gauge kinetic coupling is a function of the inflaton field with the action $\Delta {\cal L} = \frac{-f(\phi)^2}{4} F_{\mu \nu} F^{\mu \nu}$
 in which $\phi$ is the inflaton field and $F_{\mu \nu}$ is the $U(1)$ gauge field strength.
 If one chooses $f(\phi) $ such that $f(\phi) \propto a^{-2}$ then a constant source of electric field energy density is turned on at the background level and the gauge field quantum fluctuations remain scale invariant. As shown in \cite{Watanabe:2009ct} the inflationary system admits an attractor solution in which the anisotropy reaches a small but cosmologically detectable level.  Cosmological perturbation for this model in which inflaton field is a real scalar field with no charge coupling to the gauge field $A_\mu$ is studied in great details in
 \cite{Dulaney:2010sq,  Gumrukcuoglu:2010yc,  Watanabe:2010fh, Yamamoto:2012sq, Funakoshi:2012ym, Bartolo:2012sd}.

 In this work we perform the cosmological perturbation theory for the model presented in
 \cite{Emami:2010rm} in which the inflaton field is a complex scalar field charged under the
 $U(1)$ gauge field with the electric charge coupling $\e$.
 Under the Abelian Higgs mechanism, the gauge symmetry is spontaneously broken and the  gauge field acquires a dynamical mass in the form $\e^2 \rho^2 A_\mu A^\mu$ in which $\rho$ is the radial component of the complex inflaton field $\phi$. As we shall see this has interesting implications for the cosmological perturbations and in generating statistical anisotropies. Namely, as in usual Abelian Higgs mechanism, one scalar degrees of freedom is eaten by the gauge field and  the longitudinal mode of the gauge field is excited.  As a result, along with the two transverse modes of the gauge field, the longitudinal excitations will also contribute into anisotropy analysis.

The rest of paper is organized as follows. In Section \ref{model} we present our model and classify the metric and matter perturbations. In Section \ref{sec-ac} we present the second order action which will be used to calculate the anisotropic curvature perturbations power spectrum
in Section \ref{power-spec}. In Section \ref{matter-leading} we demonstrate that the leading perturbations come from the matter sector. Summary and conclusions are given in Section \ref{Summary}.
We relegate technical discussions about the choice of our gauge, integrating out non-dynamical fields and the detailed forms of the second order action into Appendices.

%%%%%%%%%%%%%%%%%%%%%%%%%%%%%%%%%%%%%%%%%%%%%%%%%%
\section{Anisotropic Inflation from Charged Scalar field }
\label{model}

Here we present our model and the metric perturbations and gauge choice in Bianchi I background.

The model we are interested in is studied at the background level in \cite{Emami:2010rm}. It contains a complex inflaton field $\phi$ which is charged under the $U(1)$ gauge field $A_\mu$ with the electric charge $\e$. The action is
\ba
 \label{action} S= \int
d^4 x  \sqrt{-g} \left [ \frac{M_P^2}{2} R - \frac{1}{2} D_\mu \phi
\,  D^\mu \bar \phi -   \frac{f^{2}(\phi)}{4} F_{\mu \nu} F^{\mu
\nu}  - V(\phi, \bar \phi) \right] \, ,
\ea
in which $M_P$ is the reduced Planck mass. To simplify the analysis, we may set $M_P=1$
but restore $M_P$ when presenting the final results for the power spectrum.

The covariant derivative is given by
\ba
D_\mu
\phi = \partial_\mu  \phi + i \e \,  \phi  \, A_\mu \, .
\ea
in which $\e$ is the electric charge coupling.

As usual, the
gauge field strength is given by
\ba F_{\mu \nu} = \nabla_\mu A_\nu
- \nabla_\nu A_\mu  = \partial_\mu A_\nu - \partial_\nu A_\mu \, .
\ea
As explained above, we have inserted the time-dependent gauge kinetic coupling $f(\phi)$ in order to break the conformal invariance such that the gauge field excitations acquire a nearly scale-invariant power spectrum and survive the exponential expansion. In order to obtain a scale-invariant gauge field power spectrum one requires $f \propto a^{-2}$. This corresponds to
a constant electric field energy density during inflation.

We assume the model is axially symmetric in field space so  $V$ and
$f(\phi)$ are only functions of $\phi \bar \phi=  |\phi |^2$. It is
more instructive to decompose the inflaton field into the radial and
angular parts \ba \phi(x) = \rho(x) \,  e^{i \theta(x)}\, , \ea so
$V=V(\rho)$ and $f^2(\phi)=f^2(\rho)$.

As usual, the action
(\ref{action}) is invariant under local gauge transformation \ba
\label{transformation} A_\mu \rightarrow A_\mu - \frac{1}{\e}
\partial_\mu \epsilon(x) \quad , \quad \theta \rightarrow \theta +
\epsilon(x) \, . \ea
In terms of $\rho$ and $\theta$  the action  (\ref{action}) is given by
\ba
 \label{action2} S= \int d^4 x \sqrt{-g} \left [ \frac{M_P^2}{2}
R -  \frac{1}{2} \partial_\mu \rho
\partial^\mu \rho-
\frac{\rho^2}{2}  \left( \partial_\mu \theta + \e A_\mu  \right)
\left( \partial^\mu \theta + \e A^\mu  \right) - \frac{f^2(\rho)}{4}
F_{\mu \nu} F^{\mu \nu}  - V(\rho) \right] \, .
\ea

We assume that the gauge field has a non-zero classical value along the $x$-direction so
$A_\mu = (0, A_x(t), 0,0)$. As a result, the background space-time is in the form of
type I Bianchi Universe with the metric
\ba
\label{bian0}
\label{metric} ds^2 = - dt^2 + e^{2\alpha(t)}\left( e^{-4\sigma(t)}d x^2
+e^{2\sigma(t)}(d y^2 +d z^2) \right) \, .
\ea
In this view $\dot \alpha$ measures the averaged Hubble expansion while $\dot \sigma(t)$ measures the level of anisotropy. In order to be consistent with cosmological observations, the level of anisotropies should be very small so $\dot \sigma/\dot \alpha \ll 1$.

The background fields equations are given in \cite{Emami:2010rm}
\ba
\label{back-A-eq}
\partial_t{\left(  f^2(\rho) e^{\alpha + 4 \sigma} \dot A_x        \right)}& =& - \e^2 \rho^2 e^{\alpha + 4 \sigma}  A_x \\
%\ddotA_{x}+\left[\dot\alpha+4\dot\sigma+2\frac{f_\rho(\rho)}{f(\rho)}\dot \rho
%\right]\dot A_{x}+e^2\rho^2f^{-2}(\rho)A_{x}&=&0 \, , \\
\label{back-rho-eq}
\ddot\rho+3\dot \alpha\dot \rho+ V_\rho+ \left(
-f(\rho)f_{,\rho}(\rho)\dot A_x^2 +\e^2 \rho A_x^2   \right) e^{-2\alpha+4\sigma}&=&0  \\
\label{Ein1-eq}
\frac{1}{2}\dot
\rho^2+V(\rho)+ \left(   \frac{1}{2}f^2(\rho)\dot
A_x^2 +\frac{\e^2\rho^2}{2}A_x^2 \right) e^{-2\alpha+4\sigma}
&=&
3 M_P^2 \left(   \dot \alpha^2-\dot \sigma^2 \right)  \\
\label{Ein2-eq}
V(\rho)+  \left(  \frac{1}{6}f^2(\rho)\dot
A_x^2+\frac{\e^2\rho^2}{3}A_x^2  \right)e^{-2\alpha+4\sigma}
&=& M_P^2 \left( \ddot \alpha    + 3 \dot \alpha^2 \right)  \\
\label{anisotropy-eq}
\left(\frac{1}{3}f^2(\rho)\dot A_x^2  -\frac{\e^2\rho^2}{3}A_x^2    \right) e^{-2\alpha+4\sigma}
&=& M_P^2\left( 3\dot \alpha \dot \sigma+ \ddot \sigma      \right)\, ,
\ea
in which a dot indicates derivative with respect to $t$.

As in conventional models of inflation, the background expansion is driven mainly by
the potential term $V$. However, the gauge field also contributes in the background expansion in the form of electric field energy density turned on along the x-direction. In order
for the anisotropy to be small we require that the electric field energy density to be very small compared to $V$. This corresponds to $R \ll 1$ in which
\ba
\label{R-def}
R \equiv \frac{\dot A_x^2 f(\rho)^2 e^{-2 \alpha}}{2 V} \, .
\ea

%%%%%%%%%%%%%%%%%%%%%%%%%%%%%%%%%%%%%%%%%%%%%%%%%%
\subsection{The Attractor Solution}

It is more convenient to express the background metric (\ref{bian0}) in the following form
\ba
\label{Bianchi-metric}
ds^2 = a(\eta)^2 (d \eta^2 + d x^2) + b(\eta)^2 ( d y^2 + d z^2)
\ea
in which $a= e^{\alpha - 2 \sigma}$ and $b=e^{\alpha + \sigma}$.
Here we have defined the conformal time $d\eta $ via $dt = a(\eta) d \eta$.
Let us define the slow-roll parameters
\ba
\epsilon_H \equiv - \frac{\dot H}{H^2}  \quad , \quad
\eta_H \equiv \epsilon_H - \frac{\ddot H}{2 H \dot H}
 \quad , \quad \dot \epsilon_H = 2 H \epsilon_H (2 \epsilon_H - \eta_H)
\ea
We are working in  the slow-roll limit in which $\epsilon_H , \eta_H \ll 1$. To leading order in slow-roll parameter and anisotropy $a \simeq b \simeq -1/H \eta$.

Although the anisotropy is very small, $R\ll 1$, so the Hubble expansion rate in modified Friedmann equation (\ref{Ein1-eq}) is mainly dominated by the isotropic potential term, but the back-reactions of the gauge field on the inflaton field induce an effective mass for the inflaton  as given by the last two terms in Eq. (\ref{back-rho-eq}). This in turn will affect the dynamics of the inflaton field. As shown in \cite{Watanabe:2009ct} the system reaches an attractor solution in which $R \propto \epsilon_H$. For this to happen we need $f(\rho) \propto a^{n}$
with $n\simeq -2$. Indeed, the background expansion is given by
\ba
\label{a-scale}
a \propto \exp \left[ - \int d \rho \frac{V}{M_P^2 V_\rho} \right] \, .
\ea
So if one chooses
\ba
\label{f-scale}
f \propto \exp \left[ -n  \int d \rho \frac{V}{M_P^2 V_\rho} \right]
\ea
this yields $f \propto a^{n}$.

The exact form of $f$ therefore depends on $V(\rho)$. For the chaotic potential used in \cite{Watanabe:2009ct}  we have
\ba
V= \frac{1}{2} m^2 \rho^2  \quad \quad  \rightarrow \quad \quad
f(\rho) = \exp {\left( \frac{c\rho^2}{2 M_P^2}  \right)}
\ea
with $c$ a constant very close to unity. As shown in \cite{Watanabe:2009ct} during the attractor phase the effective inflaton mass is reduced by the factor $1/c$ such that
during the attractor phase the inflaton evolution is given by
$d \rho/d \alpha \simeq - M_P^2V_{,\rho}/c V$. The cosmological perturbations for this background was studied in details in  \cite{Dulaney:2010sq,  Gumrukcuoglu:2010yc,  Watanabe:2010fh, Yamamoto:2012sq, Funakoshi:2012ym, Bartolo:2012sd} with the conclusion that in order not to produce too much anisotropy one needs $c-1 \sim 10^{-5}$.

For our model, following \cite{Emami:2010rm}, we consider the symmetry breaking potential which is physically well-motivated for the charged scalar field in the light of Abelian Higgs mechanism. The potential is
\ba
\label{sym-V}
V= \frac{\lambda}{4} \left( |\phi|^2 - \frac{\cM^2}{\lambda}  \right)^2
\ea
in which $\lambda$ is a dimensionless coupling. The potential has global minima at
$\mu = \pm \cM/\sqrt \lambda$. The inflaton field rolls near the top of the potential so in the slow-roll limit, the potential can be approximated by
\ba
\label{sym-V2}
V\simeq \frac{ \cM^4}{4\lambda} - \frac{\cM^2}{2} \rho^2 \, .
\ea
From Eq. (\ref{a-scale}) we have
\ba
a \propto \rho^{-p_c/2} \quad , \quad  p_c \equiv \frac{\cM^2}{2 \lambda M_P^2} \, .
\ea
%$a \propto \rho^{-p_c/2}$ in which  \ba p_c \equiv \frac{M^2}{2 \lambda M_P^2} \, . \ea
To have a long enough period of slow-roll inflation we require $p_c \gg 1$.

Motivated by this, from Eq. (\ref{f-scale}) we see that to find an attractor solution with a near scale invariant gauge field power spectrum (i.e. a scale invariant electric field power spectrum) we take \cite{Emami:2010rm}
$ f(\rho) \propto \rho^{-p} $ with $p$ very close to $p_c$.  Noting that $\rho \propto a^{-2/p_c} \propto (-\eta)^{-2/p_c}$ this yields
\ba
\label{f-eq}
f= \left( \frac{\eta}{\eta_e}\right)^{2 c}  \quad , \quad c \equiv \frac{p}{p_c} \, ,
\ea
in which $\eta_e$ indicates the time of end of inflation. We assume that at the end of inflation
$f$ reaches its canonical value $f(\tau_e) =1$ and the isotropic FRW universes emerges at the end of inflation.
As we shall see, the strength of anisotropy is measured by the parameter $I$ given by
\ba
I \equiv \frac{c-1}{c} = \frac{p-p_c}{p}\, .
\ea
During the attractor phase \cite{Watanabe:2009ct, Emami:2010rm}
\ba
R \simeq \frac{I \epsilon_H}{2}  \quad \quad , \quad \quad
\frac{\dot \sigma}{H} \simeq \frac{2R }{3}
\simeq \frac{\epsilon_H}{3}.
\ea
This indicates that the anisotropy is at the order of slow-roll parameter during the attractor phase.

At the background level there is no restriction on the value of $c$ or $I$, only one requires 
$c \ge 1$ to reach the attractor solution. However, as we shall see from the perturbation theory in next Sections, in order not to produce too much anisotropies one requires $c\rightarrow 1$ and $I \ll 1$. 

In this picture inflation ends when the back-reaction of the gauge field on the inflaton field
via the interaction $\e^2 \rho^2 A_\mu A^\mu$ induces a large mass for the inflaton.
Comparing this with the inflaton mass $\cM$, inflation ends when
$\e^2 e^{-2 \alpha_e}A_x^2 (\eta_e) \sim \cM^2$ in which $\alpha_e$ indicates the number of e-folds at the end of inflation. As shown in \cite{ Emami:2010rm} the end of inflation depends logarithmically on $\e$. More specifically, noting that during the attractor phase  \cite{ Emami:2010rm} $A_x \propto e^{3 \alpha}$, we obtain
\ba
\alpha_e \sim -\frac{\ln \e}{2} + ...
\ea
where dots indicate the dependence on other parameters such as $p_c$ and the initial value of the gauge field. As one  expects, the larger is the gauge coupling $\e$, the shorter is the period of inflation. This is easily understood from the induced mass term $\e^2 A_\mu A^\mu \rho^2$ for the inflaton field due to Higgs mechanism.

%%%%%%%%%%%%%%%%%%%%%%%%%%%%%%%%%%%%%%%%%%%%%%%%%%
\subsection{Perturbations}

Now we look at the perturbations of the background metric (\ref{Bianchi-metric}). Because the gauge field has a component along the $x$-direction, the three-dimensional rotation invariance is broken into a subset of two-dimensional rotation invariance in $y-z$ plane. Therefore, to classify our perturbations, we can look at the transformation properties of the physical fields under the rotation in $y-z$ plane.
As mentioned in \cite{Dulaney:2010sq,  Gumrukcuoglu:2010yc,  Watanabe:2010fh} the metric and matter perturbations are divided into scalar and vector perturbations for a general rotation in $y-z$ plane. It is also important to note that there are no tensor excitations in two dimensions.

The most general form of metric perturbations is
\ba
\label{deltag}
\delta g_{\alpha \beta} =   \left(
\begin{array}{c}
- 2 a^2 A~~~~~~~~~~~~  a^2 \partial_x \beta~~~~~~~~~~~~~~~~~~~~ a\, b \left( \partial_i B + B_i \right)
\\\\
   ~~~~~~~~~~~~~~~~~~~~~~~~      - 2 a^2 \bar \psi   ~~~~~~~~~~~~~~~~~~       a b\,  \partial_x \left( \partial_i \gamma + \Gamma_i \right)
\\\\
 ~~~~~~~~~~~~~~~~~~~~~~~~~~~~~~~~~~~~~~~~~~~~~~~~~~~~~~~~~~     b^2 \left( - 2 \psi \delta_{ij} + 2 E_{, ij} + E_{i,j} +E_{j,i} \right)
\end{array}
\right )  \, .
\ea
Here $A, \beta, B, \bar \psi, \gamma, \psi $ and $E$ are scalar perturbations and $B_i, \Gamma_i$ and $E_i$ are vector perturbations subject to transverse conditions
\ba
\label{transverse}
\partial_i E_i = \partial_ i B_i = \partial_i \Gamma_i =0 \, .
\ea
In Appendix \ref{appendix1} we have presented the properties of metric perturbations under a general coordinate transformation. In our analysis below we chose the following gauge
\ba
\label{gauge0}
\psi= \bar \psi = E = E_i =0 \, ,
\ea
which from Appendix \ref{appendix1}  one can check that it  is a consistent gauge.
Note that the gauge (\ref{gauge0}) is similar to the flat gauge in standard FRW background.

As for the matter sector we choose the unitary gauge $\theta=0$, so $\phi$ is real. Also, exploiting the  two-dimensional rotation symmetry, in Fourier space  we choose
\ba
\label{ab}
\overrightarrow{k}= (k_{x} , k_{y} , 0) ~~~,~~~ k_{x} = k \cos{\theta} ~~~,~~~k_{y} =\frac{b}{a} k \sin{\theta}~~.
\ea
Therefore the scalar and vector perturbations of the matter sector, $\delta A_\mu^{(S)}$  and $\delta A_\mu^{(V)}$, are
\ba
\delta A_\mu^{(S)} = (\delta A_0, \delta A_x, \partial_y M, 0) \quad \quad , \quad \quad
\delta A_\mu^{(V)} =(0, 0, 0, D) \, .
\ea
With these decompositions of the metric and matter fields into the scalar and vector sectors, one can check that these modes do not mix with each other and one can look at their excitations and propagation separately. In this work we concentrate on the anisotropies generated from scalar excitations which are more dominant compared to the anisotropies generated by vector excitations. Therefore, for the rest of analysis we set $D= \Gamma_i= B_i=0$.
%%%%%%%%%%%%%%%%%%%%%%%%%%%%%%%%%%%%%%%%%%%%%%%%%%
\subsection{Slow-roll Approximations}

In next Section we need to calculate the second order action in the slow-roll approximations.
Here we present some useful equations in the slow-roll approximation which will be employed in next section.  Including the first slow-roll and anisotropy corrections into the background expansion one can check that
\ba
\label{ab-coorections}
a \simeq  H^{-1} (-\eta)^{-1 - \epsilon_H} \quad , \quad
b \simeq  H^{-1}(-\eta)^{-1 - \epsilon_H - I \epsilon_H} \, .
\ea
Our convention is such that at the start of inflation $a_{in}=1$ with number of e-folds $N_{in}=0$. The total number of e-folds at the end of inflation is $N_e$ with $N_e\simeq 60$
to solve the flatness and the horizon problem. Furthermore, at the end of inflation
$\eta =\eta_e \rightarrow 0$. With this convention, for the CMB scale modes $k_{CMB}$, we have
$N_e = -\ln (- k_{CMB} \, \eta_e)$.
In our discussions below, we concentrate on CMB scale modes so to simplify the notation we denote $k_{CMB}$  by $k$.

From the above formulae, and using Eq. (\ref{f-eq}) for the function $f(\eta)$,
one can obtain the following expressions which would be useful later on
\begin{align}
\label{slow roll}
&&\frac{a^{'}}{a} &= (-\eta)^{-1}(1 + \epsilon_{H} )   &&\frac{b^{'}}{b} = (-\eta)^{-1}(1 + \epsilon_{H} + I\epsilon_{H})\nonumber\\
&&\frac{a^{''}}{a} &= (-\eta)^{-2}(2 + 3\epsilon_{H})  &&\frac{b^{''}}{b} = (-\eta)^{-2}(2 + 3\epsilon_{H} + 3 I\epsilon_{H})\nonumber\\
&&\frac{k^{'}}{k} &= (-\eta)^{-1}(-\sin^2{\theta}I\epsilon_{H})   &&  \frac{k^{''}}{k} = (-\eta)^{-2}(-\sin^2{\theta}I\epsilon_{H}) \nonumber\\
&& \frac{f'}{f} &= (-\eta)^{-1} ( -2 - 2 \epsilon_H - \eta_H + 2 I \epsilon_H )   &&  \frac{f^{''}}{f} = (-\eta)^{-2}( 2 + 9 \epsilon_H - 3 \eta_H + 6 I \epsilon_H   ) \, ,
\end{align}
in which a prime indicates derivative with respect to conformal time.

For the future reference, the following equations are helpful
\ba
\label{epsilon-eq}
\epsilon_H \simeq \frac{8 \lambda^2 M_P^2 \rho^2}{\cM^4} \quad , \quad
H \simeq \frac{\cM^2}{\sqrt{12 \lambda} M_P} \, .
\ea

%%%%%%%%%%%%%%%%%%%%%%%%%%%%%%%%%%%%%%%%%%%%%%%%%%
\section{Second Order Action}
\label{sec-ac}

Here we present the second order action for the scalar  perturbations. Our goal is to find the second order action both for the free fields and for the interactions. As we shall see the fields $\delta A_0, A, \beta$ and $B$ are non-dynamical in the sense that they have no time-derivatives in the action. As a result, their equations of motion give constraints which can be used to eliminate them in terms of the remaining dynamical fields
$\delta \rho, \delta A_1, M$  and $ \gamma$.

The second order action for the scalar perturbations is
\ba
\label{S2-scalar}
&&S_2^{(S)} = \int d \eta d^3 x \left[ 2 b b' A_{,x} \beta_{, x} + a b (\frac{a'}{a} +\frac{b'}{b} )
A_{,y} B{, y} +  a b \gamma_{, xy} A_{, xy} - a^2 b^2 V(\rho_0) A^2 - \frac{\e^2}{2} b^2 \rho_0^2 A_x^2 A^2 - \frac{ab}{2} \beta_{, xy} B_{, xy}
\right. \nonumber \\ &&\left.
+ a'b \gamma_{, xy} \beta_{, xy}+ \frac{ab}{2} \gamma_{, xy} \beta'_{, xy}
 + \frac{\e^2}{2} b^2 \rho_0^2 A_x^2 \beta_{x}^2 + \frac{a^2}{4} \beta_{,x y}^2  - \frac{b^2}{2}  B_{,xy}\gamma'_{, xy} + \frac{b^2}{2} (\frac{b'}{b} - \frac{a'}{a})\gamma_{, xy} B_{, xy} + \frac{b^2}{4} B_{, xy}^2 + \frac{b^2}{4}  (\gamma'_{, xy})^2 \right. \nonumber \\ &&\left.
 - \frac{\e^2}{2} b^2 A_x^2 \rho_0^2 \gamma_{, xy}^2  + \frac{b^2 f^2}{2 a^2} (A_x')^2\gamma_{, xy}^2 - \frac{b^2}{4} (\frac{b''}{b} - \frac{a''}{a}) \gamma_{, xy}^2
 + \frac{b^2}{2} \delta \rho'^2 - b^2 \rho_0' A \delta \rho' - b^2 \rho_0' \beta_{,x} \delta \rho_{,x}  - a b \rho_0' B_{, y} \delta \rho_{, y} \right. \nonumber \\ &&\left.
 - \frac{b^2}{2} \delta \rho_{,x}^2 - \frac{a^2}{2} \delta \rho_{,y}^2 + \frac{\e^2}{2} b^2 \rho_0^2 \delta A_0^2 - \e^2 b^2 \rho_0^2 A_x \beta_{, x} \delta A_0 - \frac{\e^2}{2} b^2 \rho_0^2 \delta A_1^2 - \frac{\e^2}{2} b^2 A_x^2 \delta \rho^2 - 2\e^2 b^2 \rho_0 A_x \delta \rho \delta A_1 \right. \nonumber \\ &&\left.
 + \e^2 a b \rho_0^2 A_x \gamma_{, xy} M_{, y}
 - \frac{\e^2}{2} a^2 \rho_0^2 M_{, y}^2 - \e^2 b^2 \rho_0^2 A_x A \delta A_1
 -\e^2 b^2 \rho_0 A_x^2 A \delta \rho + \frac{f^2 b^2}{2 a^2} \delta A_1'^2
 +\frac{f^2 b^2}{2 a^2} \delta A_{0 , x}^2  \right. \nonumber \\ &&\left.
 - \frac{f^2 b^2}{ a^2} \delta A_1' \delta A_{0, x} -  \frac{f^2 b^2}{ a^2} A_x' A \delta A_1'
 +  \frac{f^2 b^2}{ a^2} A_x' A \delta A_{0, x}  - \frac{f^2 b}{ a} A_x' \gamma_{, xy} M_{,y}'
 +\frac{f^2 b}{ a}  A_x'  \gamma_{, xy} \delta A_{0, y} - \frac{f^2 b}{ a}  A_x' B_{, y} \delta A_{1, y} \right. \nonumber \\ &&\left.
 + \frac{f^2 b}{ a} A_x' B_{, y} M_{, xy} + \frac{f^2}{2} M_{y}'^2 + \frac{f^2}{2} \delta A_{0, y}^2 - f^2 M_{, y}' \delta A_{0, y} - \frac{f^2}{2} \delta A_{1, y}^2 -  \frac{f^2}{2}
 M_{, xy}^2 + f^2 \delta A_{1, y} M_{, xy} + 2 \frac{f f_{,\rho} b^2}{a^2} A_x'\delta A_1' \delta \rho
  \right. \nonumber \\ &&\left.
  - 2 \frac{f f_{,\rho} b^2}{a^2} A_x' \delta A_{0, x} \delta \rho - \frac{f f_{,\rho} b^2}{a^2} A \delta \rho +  \frac{ f_{,\rho}^2 b^2}{2a^2}A_x'^2 \delta \rho^2 + \frac{f f_{,\rho \rho} b^2}{2a^2} A_x'^2 \delta \rho^2 - \frac{a^2 b^2}{2} V_{, \rho \rho} \delta \rho^2 - a^2 b^2 V_{, \rho} A \delta \rho ~~~
 \right]
\ea
in which a prime indicates derivative with respect to conformal time.

As mentioned above, the excitations $\delta A_0, A, \beta$ and $B$ have no time-derivatives so they are non-dynamical. The details of eliminating the non-dynamical excitations in terms of
dynamical perturbations are given in Appendix \ref{reduced}.

The final second order action is a complicated function of $\delta \rho, \delta A_1, M$  and $ \gamma$. Specifically, integrating out  $\delta A_0, A, \beta$ and $B$ one encounters the functions $\lambda_i$ and
$\bar \lambda_i$ as defined in Eqs. (\ref{lambdak1}) - (\ref{lambda17}). At this level it seems hopeless to get any insight into the form of the action and the prospects for analytical analysis. Happily, the analysis becomes considerably simple if one notice the following effects. Looking at the formulae for $\lambda_i$ and $\bar \lambda_i$ it is evident that  $\bar \lambda_1$ is the  key parameter which controls the form of other $\lambda_i$ and $\bar \lambda_i$.  Now let us look at the function $\bar\lambda_1$
\ba
\label{b-lambda1}
\bar \lambda_1 \equiv \frac{b^2}{2 a^2} k^2 f^2 + \frac{\e^2 }{2} b^2 \rho^2 \, .
\ea
Following the procedures of integrating out the non-dynamical fields in Appendix \ref{reduced} one can check that $\bar \lambda_1$ comes from integrating out $\delta A_0$.
Neglecting the anisotropy for the moment, the ratio of the second term in $\bar \lambda_1$
compared to the first term scales like $\e^2 a^2/ f^2 \sim \e^2 H^2  \eta_e^4/\eta^6 $. Therefore, during the early stages of inflation in which $-\eta \gg -\eta_e$, the second term in $\bar \lambda_1$ is completely negligible compared to the first term. In this limits all $\lambda_i$
and $\bar \lambda_i$ collapse to simple forms and we will be in the limit somewhat similar to
\cite{Watanabe:2010fh}. In this limit the effect of gauge coupling $\e$ is sub-dominant in the action and the leading interaction comes from the gauge kinetic coupling $f^2(\phi) F^2$.
%as in  \cite{Watanabe:2010fh},  the mode excitations are massless.
On the other hand, as inflation proceeds the second term in
$\bar \lambda_1$ eventually dominates and we enter the second phase in which all
$\lambda_i$ and $\bar \lambda_i$ are proportional to $\e^2$. In this limit, the interaction induced from the symmetry braking, $\e^2 \rho^2 A_\mu A^\mu$, becomes as important as the interaction from the gauge kinetic coupling.   We will elaborate more on this issue later on when we present the dominant interactions for the transverse and longitudinal modes.

Having this said, one may wonder why $\bar \lambda_1$ plays such a prominent role.
The answer to this question is provided in Section \ref{matter-leading} in which we demonstrate that the leading interactions come from the matter sector. So it is not surprising that only $\bar \lambda_1$, which originates from integrating out $\delta A_0$, will have a prominent effect
while the other parameters $\lambda_i$ and $\bar \lambda_i$, which have their origins in integrating out metric fields $A, \beta$ and $B$, are negligible.

The time when the two terms in $\bar \lambda_1$ become comparable, denoted by $\eta_c$, is given by
\ba
\label{etac}
-\eta_c =  \left(\frac{\e \rho}{-H k }\right)^{1/3} (-\eta_{e})^{2/3}=
\left( \frac{3 \e M_{P} \sqrt{2\epsilon_H}}{\cM^2} \right)^{1/3} (-\eta_e)^{2/3} \, .
\ea
In order to obtain the last equality, we considered  the CMB scale modes in which
$k= a_{in} H = H$.  Eq. (\ref{etac}) indicates a $k$-dependence in $\eta_c$. However, as we will see explicitly below, the leading contributions from the inflaton field and the transverse mode are blind to
this $k$-dependence.

It is also instructive to look at $N_c$, the number of e-folds when $\eta=\eta_c$. Using $\eta \simeq -1/a H$ and Eq. (\ref{etac}) we have
\ba
\label{Nc}
N_c \simeq \frac{2 N_e}{3} - \frac{1}{3} \ln \left( \e \sqrt{\frac{3 \epsilon_H}{2 \lambda}}
\right) \simeq \frac{2 N_e}{3} \, .
\ea
The last approximation is valid for typical parameter values such that the logarithmic correction
in Eq. (\ref{Nc}) is at the order of unity. Our convention is such that at the start of inflation
$N_{in}=0$ and  the total number of e-folds at the end of inflation is $N_e$.  With $N_e\simeq 60$ to solve the flatness and the horizon problem we obtain  $N_c \sim 40$.

%%%%%%%%%%%%%%%%%%%%%%%%%%%%%%%%%%%%%%%%%%%%%%%%%%
\subsection{Second Order Action in the Slow-roll Approximation}
%\subsubsection{Slow roll approximation}
After integrating out the non-dynamical fields, the remaining dynamical fields are $\delta \rho,  \delta A_1, M$ and $\gamma$. However, for the gauge field excitations, the physically relevant fields are the transverse mode $D_1$ and the longitudinal mode $D_2$ which are related to $\delta A_1$ and $M$ via
\ba
\label{D12}
D_{1} &\equiv& \delta A_{1} -ik \cos{\theta}M \\
D_{2}&\equiv& \cos{\theta} \delta A_{1} + ik \sin^2{\theta}M \, .
\ea
Here we present the second order action in the slow-roll limit for the dynamical variables
$\delta \rho,  D_1$ and  $D_2$.  The action is presented separately for $\eta< \eta_c$ and $\eta_c < \eta < \eta_e$. 

In this work we are interested in anisotropy generated in curvature perturbation power spectrum. Note that,  as discussed in Appendix \ref{appendix1}, the scalar perturbations 
$\gamma$ will furnish one polarization of tensor perturbations in isotropic universe after inflation. Therefore the interactions $L_{\delta \rho \gamma}, L_{D_1 \gamma}$ and $L_{D_2 \gamma}$ will not contribute into curvature perturbation anisotropy and we do not present them in this section. However, they are presented in the Appendices when we present the whole second order action for the scalar perturbation.

%%%%%%%%%%%%%%%%%%%%%%%%%%%%%%%%%%%%%%%%%%%%%%%%%%
\subsection{ $\eta  < \eta_c $ }
\label{eta-less}
%\paragraph{\textbf{First Phase :}\\}
First we consider the period in which $\eta < \eta_c$ so the term containing $\e$ in
$\lambda_i$ and $\bar \lambda_i$ are negligible and the first term in $\bar \lambda_1$ in Eq. (\ref{b-lambda1}) dominates.

As we shall see in next section, in order not to produce too much anisotropy, one requires $I \ll 1$ (i.e. c $\rightarrow 1$) which we will assume in all our analysis below.  Considering the leading corrections from the slow-roll and anisotropy expansion yields (for details see Appendix \ref{second slow})
\begin{align}
\label{Total scalar action}
S_{2}^{(1)} = \int d\eta \,  d^3k \bigg{(} L_{\rho \rho}  + L_{D_{1}D_{1}} + L_{D_{2}D_{2}} +    L_{\rho D_{1}} +  L_{\rho D_{2}} +     L_{ D_{1} D_{2}} \bigg{)}~,
\end{align}
in which the free fields Lagrangians are
\ba
\label{rhorho action1}
L_{\rho \rho} &=& \frac{1}{2}|\overline{\delta \rho}'|^2 + \frac{1}{2}\bigg{[} -k^2 + (-\eta)^{-2}\bigg{(}2+9\epsilon_{H}-6\frac{\eta_{H}}{1-I}
-12\frac{I}{1-I}(1-2\sin^2{\theta})\bigg{)}\bigg{]}|\overline{\delta \rho}|^2
\\
\label{D1D1  action1}
L_{D_{1} D_{1}} &=& \frac{1}{2}|\overline{D_{1}}^{'}|^2 + \frac{1}{2}\bigg{[} -k^2 + (-\eta)^{-2}\bigg{(}2+9\epsilon_{H}-3\frac{\eta_{H}}{1-I}\bigg{)}\bigg{]}|\overline{D_{1}}|^2
\\
\label{D2D2  action1}
L_{D_{2} D_{2}} &=& \frac{1}{2}|\overline{D_{2}}^{'}|^2 + \frac{1}{2}\bigg{[} -k^2 + (-\eta)^{-2}\bigg{(}2+3\epsilon_{H} + I\epsilon_{H}\bigg{)}\bigg{]}|\overline{D_{2}}|^2 \, .
\ea
Here we have defined the canonically normalized fields via
\begin{align}
\label{canonical variables1}
\overline{\delta \rho}_{k} &\equiv b \delta \rho_{k}\equiv u_{k} \\
\label{D1bar}
\overline{D_{1k}}&\equiv \frac{b}{a}f \sin{\theta} D_{1k}\equiv \frac{b}{a}f \sin{\theta} v_{k}\\
\label{D2bar}
\overline{D_{2k}}&\equiv \frac{\e \cM^2}{2\sqrt{2}\lambda k M_{P}}\sqrt{\frac{\epsilon_{H}}{1-I}} bD_{2k} \equiv \frac{\e \cM^2}{2\sqrt{2}\lambda k M_{P}}\sqrt{\frac{\epsilon_{H}}{1-I}} b w_{k} 
\end{align}

The interaction Lagrangians relevant for curvature perturbations anisotropy are
\ba
\label{rhoD1new}
L_{\rho D_{1}} &=& \left(\frac{1}{\eta}\right)\frac{b^2}{a} \sqrt{6I} \sin^2{\theta} f \Big{(} \delta \rho^{*} D_{1}^{'} + c.c.\Big{)} - \left( \frac{a^2}{f\eta} \right) \e^2 \sqrt{\frac{I \epsilon_{H}^2}{\lambda}}M_{P} \sin^2{\theta}\Big{(} \delta \rho^{*} D_{1} + c.c.\Big{)}
\\
\label{rhoD2new}
L_{\rho D_{2}} &=& \left(\frac{1}{\eta}\right) \left(\frac{ab^2}{8}\frac{\e^2 \cM^4}{\lambda^2 k^2 f}\right)\sqrt{6I} \epsilon_{H} \cos^3{\theta}\Big{(} \delta \rho^{*} D_{2}^{'} + c.c.\Big{)} - \left( \frac{a^2}{f\eta} \right) \e^2 \sqrt{\frac{I \epsilon_{H}^2}{\lambda}}M_{P} \cos{\theta}\Big{(} \delta \rho^{*} D_{2} + c.c.\Big{)}
\ea

As mentioned  before, Eqs. (\ref{rhorho action1})-(\ref{D2D2  action1}) represent the free-field actions for  $\bar \delta \rho,  \bar D_1$ and $\bar D_2$. As expected, during this phase in which the effect of symmetry breaking term $\e^2 \rho^2 A_\mu A^\mu$ is sub-leading, similar to \cite{Watanabe:2010fh}, Eqs. (\ref{rhorho action1})-(\ref{D2D2  action1}) represent nearly massless fields with almost scale-invariant power spectrum.
The interaction terms are given by Eqs. (\ref{rhoD1new}) and (\ref{rhoD2new}). For technical reasons the interaction terms are presented in terms of the original non-canonical fields. 

To calculate the induced anisotropy in curvature perturbation power spectrum, we are interested in interactions between the gauge field and the inflaton field given by $L_{\rho D_1}$ and $L_{\rho D_2}$ in Eqs. (\ref{rhoD1new}) and (\ref{rhoD2new}). First let us look at the interaction between the transverse mode and the inflaton field, $L_{\rho D_1}$.  From Eq. (\ref{rhoD1new}) we see that
$L_{\rho D_1}$ has two contributions. The first term in  $L_{\rho D_1}$ comes from the gauge kinetic coupling $f^2(\phi) F^2$ which is similar to models such as
\cite{Watanabe:2010fh} with a real inflaton field. However, the second term in $L_{\rho D_1}$
comes from the interaction $\e^2 \rho^2 A_\mu A^\mu$ which originates  from the symmetry breaking effects. This interaction does not exist in models where $\phi$ is a real field. One  can easily check that for $\eta < \eta_c$  the first term in $L_{\rho D_1}$ dominates over the second term. The two interactions in $L_{\rho D_1}$
 become comparable near $\eta = \eta_c$. This is understandable, since during the period $\eta< \eta_c$, the effects of symmetry breaking are small and the system proceeds as in \cite{Watanabe:2010fh}.

Now let us look at  $L_{\rho D_2}$,  the interaction between the longitudinal mode and the inflaton field.  As expected the longitudinal mode becomes physical because of the symmetry breaking effect $\e^2 \rho^2 A_\mu A^\mu$ so both terms in Eq. (\ref{rhoD2new}) are proportional to $\e^2$. The last term in $L_{\rho D_2}$ comes directly from the interaction
$\e^2 \rho^2 A_\mu A^\mu$. However, the first term in  $L_{\rho D_2}$ is somewhat non-trivial. As we shall see in Section \ref{matter-leading}, after integrating out $\delta A_0$ a coupling in the form  $\delta \rho^*D_2'+ c.c.$ appears which
cancels the corresponding term coming from  $f^2 F^2$ interaction during the phase $\eta < \eta_c$. As a result, the derivative coupling $\delta \rho^*D_2'+ c.c.$ during the first phase comes from sub-leading interactions so it contains $\e^2$. Finally, comparing the two terms  in Eq. (\ref{rhoD2new})
one can check that during the phase $\eta< \eta_c$ the second term in Eq. (\ref{rhoD2new})
is smaller than the first term by a factor $1/p_c \ll 1$.

It is also instructive to compare $L_{\rho D_1}$ and $L_{\rho D_2}$ during this phase. Relating
$D_1$ and $D_2$  to the normalized field $\bar D_1$ and $\bar D_2$ as given in Eqs. (\ref{D1bar}) and (\ref{D2bar}) and assuming that $\bar D_1$ and $\bar D_2$ have similar amplitudes one can check that
\ba
\label{L-ratio}
\frac{L_{\rho D_1}}{L_{\rho D_2}} \sim \frac{k\,  f}{ \e \rho a} \gg1
\ea
in which Eq. (\ref{epsilon-eq}) have been used to eliminate $\epsilon_H$. The conclusion that
$L_{\rho D_1} \gg L_{\rho D_2}$ is understandable since during the first phase the effects of the
coupling $\e$ is negligible.

To summarize, the leading interaction during the phase $\eta < \eta_c$ is given by the first term in Eq.(\ref{rhoD1new}) from the transverse mode interaction $L_{\rho D_1}$. As mentioned, this interaction originates from the gauge kinetic coupling interaction $f(\rho)^2 F^2$. As a result, the induced anisotropy originated from this phase is similar to models with a real inflaton field such as in  \cite{Watanabe:2010fh}.

%%%%%%%%%%%%%%%%%%%%%%%%%%%%%%%%%%%%%%%%%%%%%%%%%%
\subsection{ $\eta_c< \eta  < \eta_e $ }
\label{eta-big}
As we mentioned below Eq. (\ref{b-lambda1}) during the period
$\eta_c< \eta  < \eta_e  $ the effect of the gauge coupling $\e$ becomes important.  During this phase the dominant contributions  in $\lambda_i$ and $\bar \lambda_i$ in Eqs. (\ref{lambdak1}) - (\ref{lambda17}) come from the terms containing $\e$. Expanding to leading order in terms of the slow-roll parameters and $I$ and concentrating on CMB-scale modes which are expected to be super-horizon by the time $\eta=\eta_c$,
the second order action is
\begin{align}
\label{Total scalar action2}
S_{2}^{(2)} = \int d\eta d^3k \bigg{(} L_{\rho \rho} +  L_{D_{1}D_{1}} + L_{D_{2}D_{2}} +   L_{\rho D_{1}} +  L_{\rho D_{2}} \bigg{)}~,
\end{align}
where,
\ba
\label{rhorho action}
L_{\rho \rho} &=& \frac{1}{2}|\overline{\delta \rho}'|^2 + \bigg{[} \frac{1}{\eta^{2}} -\left(\frac{\e^2 I \epsilon_{H} \lambda }{\cM^4}\right) \left(\frac{1}{f^{2}\eta^{2}}\right)\bigg{]}|\overline{\delta \rho} |^2
\\
\label{D1D1  action}
L_{D_{1} D_{1}} &=& \frac{1}{2}| \overline{D_{1}}'|^2 + \bigg{[} \frac{1}{\eta^{2}} -\left(\frac{3\e^2 \epsilon_{H}}{4\lambda}\right) \left(\frac{1}{f^{2}\eta^{2}}\right)\bigg{]}| \overline{D_{1}} |^2
\\
\label{D2D2  action}
L_{D_{2} D_{2}} &=& \frac{1}{2}| \overline{D_{2}}'|^2 + \bigg{[} \frac{1}{\eta^{2}} -\left(\frac{3\e^2 \epsilon_{H}}{4\lambda}\right) \left(\frac{1}{f^{2}\eta^{2}}\right)\bigg{]}| \overline{D_{2}} |^2
\\
\label{rhoD1 action20}
L_{\rho D_{1}} &=& \left(\frac{1}{\eta}\right)\frac{b^2}{a} \sqrt{6I} \sin^2{\theta} f \Big{(} \delta \rho^{*} D_{1}^{'} + c.c.\Big{)} - \left( \frac{a^2}{f\eta} \right) \e^2 \sqrt{\frac{I \epsilon_{H}^2}{\lambda}}M_{P} \sin^2{\theta}\Big{(} \delta \rho^{*} D_{1} + c.c.\Big{)}
\\
\label{rhoD2 action20}
L_{\rho D_{2}} &=& \left(\frac{1}{\eta}\right)\frac{b^2}{a} \sqrt{6I} \cos{\theta} f \Big{(} \delta \rho^{*} D_{2}^{'} + c.c.\Big{)} - \left( \frac{a^2}{f\eta} \right) \e^2 \sqrt{\frac{I \epsilon_{H}^2}{\lambda}}M_{P} \cos{\theta}\Big{(} \delta \rho^{*} D_{2} + c.c.\Big{)} \, .
\ea

During the second phase  the canonical variables $\overline{\delta \rho}_{k}$ and $\overline{ D_1}_{k}$ are the same as defined in Eqs. (\ref{canonical variables1}) and (\ref{D1bar}) while the canonical normalized field $\overline{D_2}_k$ is
\ba
\label{canonical variables2}
%\overline{\delta \rho}_{k} &\equiv& b \delta \rho_{k}\equiv u^{(m)}_{k} \\
%\overline{D_{1k}}&\equiv& \frac{b}{a} \sin{\theta} f D_{1k} \equiv \frac{b}{a} \sin{\theta} f v^{(m)}_{k} \\
\overline{D_{2k}}&\equiv& \frac{b}{a} f D_{2k}\equiv \frac{b}{a} f w_{k} \, .
\ea

As in the first phase, for the purpose of  calculating the curvature perturbations power spectrum, we look into interactions  between $\delta \rho$ and other fields. As before, the interaction  $L_{\rho \gamma}$ does not have any directional dependence so we have not considered it in above action. Therefore we are left with $L_{\rho D_{1}}$ and $L_{\rho D_{2}}$.

The crucial difference compared to the first phase is that once the second term in Eq. (\ref{b-lambda1}) dominates over the first term, the effects of gauge coupling $\e$ from the interaction $\e^2 A_\mu A^\mu$ become important. To see this, let us look at the interactions $L_{\rho D_1}$ and $L_{\rho D_2}$ given in Eqs. (\ref{rhoD1 action20}) and (\ref{rhoD2 action20}). One can easily check that in both Eqs. (\ref{rhoD1 action20}) and (\ref{rhoD2 action20}), the terms containing $\e^2$ are much larger than the first terms containing $D_1'$ and $D_2'$ which
come from the gauge kinetic coupling $f^2 F^2$. In this view, during $\eta_c < \eta< \eta_e $ the dominant interaction in the system is $\e^2 A_\mu A^\mu$ and not $f^2 F^2$. This is in contrast to the first phase in which, as we saw in the previous subsection,  the interaction $f^2 F^2$ was the dominant one and the effects of symmetry breaking were not important.

It is also instructive to compare the forms of $L_{\rho D_{1}}$ and $L_{\rho D_{2}}$
for these two phases. From Eq. (\ref{rhoD1 action20}) and (\ref{rhoD1new}) we see that
$L_{\rho D_{1}}$ has the same functional form in both phases. However, $L_{\rho D_{2}} $
has different functional forms in two phases. The last terms in Eq. (\ref{rhoD2 action20})
and (\ref{rhoD2new}) are the same. This is reasonable since this term directly originates from
the interaction $\e^2 A_\mu A^\mu$. However, the first terms in Eq. (\ref{rhoD2 action20})
and (\ref{rhoD2new}),  containing the derivative coupling of $\delta \rho^*D_2'+ c.c.$,  have different forms in these two phases. Intuitively, this is somewhat non-trivial. However, as we shall show explicitly in Section  \ref{matter-leading}, this difference originates from integrating out $\delta A_0$.
After integrating out $\delta A_0$, a coupling in the form  $\delta \rho^*D_2'+ c.c.$ appears which
cancels the corresponding term coming from  $f^2 F^2$ interaction in the first phase. As a result, the derivative coupling $\delta \rho^*D_2'+ c.c.$ during the first phase comes from sub-leading interactions so it contains $\e^2$. However, during the second phase, the leading terms
in derivative coupling $\delta \rho^*D_2'+ c.c.$ survives and as a result the first term in
Eq. (\ref{rhoD2 action20}) gets the usual form similar to derivative coupling in Eq. (\ref{rhoD1 action20}).

Comparing Eq. (\ref{canonical variables2}) with Eq. (\ref{D2bar}) we see that
$\overline {\delta \rho}$ and $\overline D_1$ have the same forms in both phases but
$\overline D_2$ have different forms in two phases. Also Eq. (\ref{rhoD1 action20}) is proportional to $\sin^2{\theta}$ while Eq. (\ref{rhoD2 action20}) is proportional to $\cos{\theta}$. As a result we can guess that the contributions of the longitudinal mode in $g_*$ has a  different sign than the corresponding contributions from the transverse mode.  So the question arises whether or not we can produce a positive $g_*$ factor from the longitudinal mode (from \cite{Watanabe:2010fh} we know that $g_{*}$ is negative for the transverse modes). We will come back to this question when we calculate the power spectrum of curvature perturbations.

Having obtained the quadratic action we also need to know the wave function solution for
$\delta \rho, D_1$ and $D_2$. For the first phase the answer is simple: since all modes are nearly massless, the mode functions of $\overline{\delta \rho}, \overline D_1$ and $\overline D_2$ are simply the mode function of the massless scalar fields with the Bunch-Davies initial condition. More specifically
\begin{align}
\label{mode function0}
M_{j\textbf{k}} &= m_{j\textbf{k}}a_{j\textbf{k}} + m^{*}_{j(-\textbf{k})}a^{\dag}_{j(-\textbf{k})} ~~~,~~~ j = (\overline{\delta \rho} , \overline{\gamma}  , \overline{D_{1}} , \overline{D_{2}})  \nonumber \\
m_{j\textbf{k}} &\equiv \frac{1}{\sqrt{2k}}e^{-ik\eta}(1-\frac{i}{k\eta}) \, .
\end{align}
The profile of the outgoing solution for $\eta_c <\eta < \eta_e$ is given in details in
Appendix \ref{outgoing modes}. Here
we demonstrate that during the second phase the inflaton excitations and the gauge field excitations remain nearly massless so one can still use the free wave function given in Eq. (\ref{mode function0}). To verify that the perturbations remain nearly massless during the second phase  it is instructive to look at the times when the arguments of the Hankel functions  Eqs. (\ref{mode function-u}),  (\ref{mode function-v}) and (\ref{mode function-w}) becomes the order unity. This can be interpret as the times when the modes become massive so it oscillates towards the end of inflation.  Defining $\eta_u$ as the time when the inflaton field fluctuations $u_k$ become massive we have $\eta_u \simeq \Omega^{1/4}$. As a result, the number of e-folds towards the end of inflation when $u_k$ is massive, $\Delta N_u \equiv \ln (\eta_u/\eta_e)$, is given by
\ba
\label{Nu}
\Delta N_u \simeq \frac{1}{4} \ln \left( \frac{\e^2 I \epsilon_H \lambda M_P^4}{\cM^4}
\right) \simeq   \frac{1}{4}  \ln \left( \frac{\e^2 I \epsilon_H }{\lambda p_c^2} \right)
\simeq \frac{1}{4}\ln(10^3 \e^2) \, ,
\ea
in which in the last approximation we assumed the typical model parameters of symmetry breaking inflation $\lambda \sim 10^{-13},  \epsilon_H \sim 10^{-2}$, $p_c \sim N_e$, and as we shall see below, $I \sim 10^{-5}$. Therefore, if $\e \lesssim 1$ which is a natural choice, we see that $\Delta N_u \sim 2$. As a result, for $\e$ not exponentially large, the inflaton field excitations remain nearly massless almost during entire period of  inflation. As a result, in our analysis of power spectrum in next section we can treat $u_k$ as nearly massless field excitations.

Also one can  check that $\Delta /\Omega \sim p_c^2/I$. As a result, the time
$\eta_v \simeq  \Delta^{1/4}$  when the gauge field excitations become massive,
and the corresponding number of e-foldings $\Delta N_v \equiv \ln (\eta_v/\eta_e)$, is given by
\ba
\Delta N_v \simeq \frac{1}{4} \ln \left(  \frac{\e^2 \epsilon_H}{\lambda} \right)
\simeq \frac{1}{4} \ln( 10^{10} \e^2) \simeq 6 + \frac{1}{2} \ln \e \, .
\ea
This indicates that   for typical model parameters $\Delta N_v - \Delta N_u \simeq 4$ so
$\Delta N_v \sim 6$. Therefore, we can also safely conclude that the gauge field excitations are
nearly massless during most of the period of inflation. Finally, one can also easily check that
$\eta_c \gg \eta_u, \eta_v$, so at the time $\eta =\eta_c$, all fields excitations are nearly massless to very good approximations.

%%%%%%%%%%%%%%%%%%%%%%%%%%%%%%%%%%%%%%%%%%%%%%%%%
\section{Power Spectrum of  Curvature Perturbations}
\label{power-spec}

We are ready to calculate the curvature perturbation power spectrum.  We are interested in anisotropies generated in curvature perturbation power spectrum. The anisotropies are generated by interactions $L_{\rho D_1}$ and $L_{\rho D_2}$ from the coupling of the transverse and longitudinal modes to $\delta \rho$. The corresponding Feynman diagrams are
given in Fig. 1.

%%%%%%%%%%%%%%%%%%%%%%%%%%%%%%%%%%%%%%%%%%%%%%%%%%
\begin{figure}
\includegraphics[ width=0.9\linewidth]{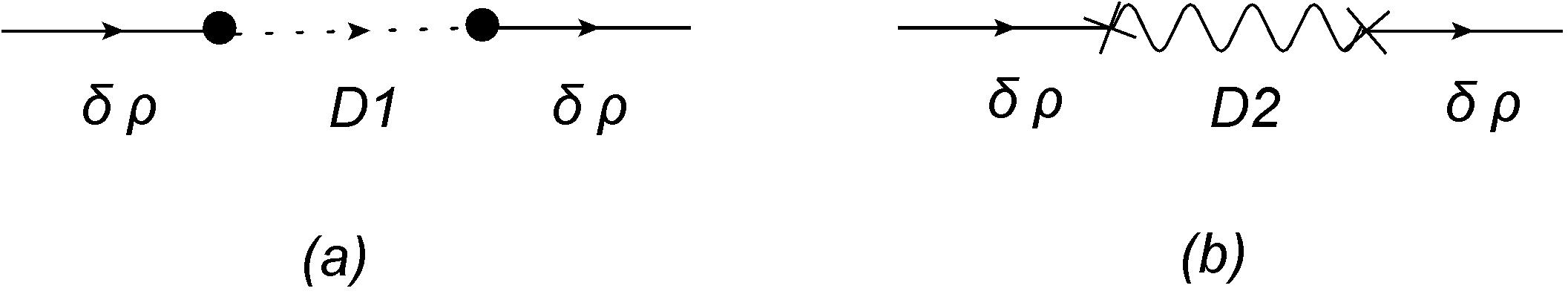}
\caption{ The transfer vertices for the interactions of the inflaton field $\delta \rho$ with the gauge field excitations $D_1$ and $D_2$. The left figure represents $L_{\rho D_1}$ as given by Eq. (\ref{rhoD1 action20})
while the right figure represents $L_{\rho D_2}$ given by Eq. (\ref{rhoD2 action20}). }
\label{stop}
\end{figure}
%%%%%%%%%%%%%%%%%%%%%%%%%%%%%%%%%%%%%%%%%%%%%%%%%%
Using the standard In-In formalism for the curvature perturbation power spectrum
\cite{Weinberg:2005vy, Chen:2010xka, Chen:2009zp} we have
\begin{align}
\label{n point function}
<\overline{\delta \rho}^2(\eta_{e})> = \bigg{<} \bigg{|}\bigg{[} \overline{T} \exp{\left(i\int_{\eta_{0}}^{\eta_{e}}H_{I}(\eta')d\eta'\right)}\bigg{]} \overline{\delta \rho}^2(\eta)\bigg{[} T \exp{\left(-i\int_{\eta_{0}}^{\eta_{e}}H_{I}(\eta')d\eta'\right)}\bigg{]}  \bigg{|}\bigg{>} \, ,
\end{align}
where $T$ and $\overline{T}$ respectively denote the time-ordered and anti-time-ordered products and $H_{I}$ refers to the interaction part of the Hamiltonian in the interaction picture.
As for $\eta_0$ we can take $\eta_0 \rightarrow -\infty$ so the modes of interests were originally deep inside the horizon.

To leading order the contribution of anisotropy in inflaton power spectrum, $\Delta <\overline{\delta \rho}^2(\eta_{e})>$, is
\begin{align}
\label{commutator}
\Delta <\overline{\delta \rho}^2(\eta_{e})> = - \int_{\eta_{0}}^{\eta_{e}} d\eta_{1} \int_{\eta_{0}}^{\eta_{1}}d\eta_{2} \bigg{[} H_{I}(\eta_{2}) , \bigg{[} H_{I}(\eta_{1}) , \overline{\delta \rho}^2(\eta)\bigg{]}\bigg{]} \, .
\end{align}

As discussed in details in previous Section the derivative interactions of the  longitudinal mode,    terms containing $D_2'$,  have different forms in  phases  $\eta < \eta_c$ and $\eta > \eta_c$. To take this into account, we  can write the interaction Hamiltonian as follows
\ba
\label{Hamiltonian}
H_{I}(\eta) &=&  -\left(\frac{1}{\eta}\right)\left(\frac{b^2f}{a}\sqrt{6I}\sin^2{\theta}\right)\left(\delta \rho ^{*} D_{1}'+c.c. \right) + \left( \frac{a^2}{f\eta} \right) \e^2 \sqrt{\frac{I \epsilon_{H}^2}{\lambda}}M_{P} \sin^2{\theta}\Big{(} \delta \rho^{*} D_{1} + c.c.\Big{)}\nonumber\\
&& +  \left( \frac{a^2}{f\eta} \right) \e^2 \sqrt{\frac{I \epsilon_{H}^2}{\lambda}}M_{P} \cos{\theta}\Big{(} \delta \rho^{*} D_{2} + c.c.\Big{)}
-\left(\frac{1}{\eta}\right)\left(\frac{b^2f}{a}\sqrt{6I}\cos{\theta}\right)\left(\delta \rho ^{*} D_{2}'+c.c. \right)\theta(\eta-\eta_{c}) \nonumber\\
&& \equiv H_{1} + H_{2} + H_{3} + H_{4} \,  ,
\ea
in which the form of interactions $H_i, i=1,..4$, is read off in order from the above equation.
Here we used the step function $\theta (\eta - \eta_c)$ to take into account the change in the form of interaction after $\eta > \eta_c$ for the longitudinal mode.

Plugging back Eq. (\ref{Hamiltonian}) into the Eq. (\ref{commutator})  the non-zero terms are
\begin{align}
\label{Correction Powerspectrum}
\frac{\Delta <\overline{\delta \rho}^2(\eta_{e})>}{<\overline{\delta \rho}^2(\eta_{e})>}
&= \frac{\Delta <\overline{\delta \rho}^2(\eta_{e})>_{11}}{<\overline{\delta \rho}^2(\eta_{e})>} + \frac{\Delta <\overline{\delta \rho}^2(\eta_{e})>_{12}}{<\overline{\delta \rho}^2(\eta_{e})>} + \frac{\Delta <\overline{\delta \rho}^2(\eta_{e})>_{21}}{<\overline{\delta \rho}^2(\eta_{e})>} + \frac{\Delta <\overline{\delta \rho}^2(\eta_{e})>_{22}}{<\overline{\delta \rho}^2(\eta_{e})>} \nonumber \\
& ~~~ + \frac{\Delta <\overline{\delta \rho}^2(\eta_{e})>_{33}}{<\overline{\delta \rho}^2(\eta_{e})>} + \frac{\Delta <\overline{\delta \rho}^2(\eta_{e})>_{34}}{<\overline{\delta \rho}^2(\eta_{e})>} + \frac{\Delta <\overline{\delta \rho}^2(\eta_{e})>_{43}}{<\overline{\delta \rho}^2(\eta_{e})>} + \frac{\Delta <\overline{\delta \rho}^2(\eta_{e})>_{44}}{<\overline{\delta \rho}^2(\eta_{e})>}
\end{align}
In this notation, $\Delta <\overline{\delta \rho}^2(\eta_{e})>_{ij}$ represents the contribution of the two interactions from $H_i$ and $H_j$ in Eq. (\ref{Hamiltonian}).

Now we calculate each term in Eq. (\ref{Correction Powerspectrum}) in turn.  The contributions from the transverse mode $D_1$ (and $D_1'$) are
\begin{align}
\label{Correction Powerspectrum11}
\frac{\Delta <\overline{\delta \rho}^2(\eta_{e})>_{11}}{<\overline{\delta \rho}^2(\eta_{e})>}
&= \frac{192 I}{|u^{(0)}(\eta_{e})|^2}
\int_{\eta_{0}}^{\eta_{e}} d\eta_{1} \int_{\eta_{0}}^{\eta_{1}}d\eta_{2} \left(\frac{\eta_{1}\eta_{2}}{\eta_{e}^4}\right)\bigg{(}\sin^4{\theta} Im \bigg{[}u(\eta_{1})u^{*}(\eta_{e})\bigg{]} Im \bigg{[}u(\eta_{2})u^{*}(\eta_{e})v^{*'}(\eta_{1})v^{'}(\eta_{2})\bigg{]}\bigg{)} \nonumber\\
& ~~~~~ = 24 I \sin^2{\theta} N_{e}^2
\end{align}
\begin{align}
\label{Correction Powerspectrum12}
\frac{\Delta <\overline{\delta \rho}^2(\eta_{e})>_{12}}{<\overline{\delta \rho}^2(\eta_{e})>}
&= \frac{32 \e^2 I \epsilon_{H}\sqrt{6}}{|u^{(0)}(\eta_{e})|^2 \sqrt{\lambda}}\frac{M_{P}}{H}
\int_{\eta_{0}}^{\eta_{e}} d\eta_{1} \int_{\eta_{0}}^{\eta_{1}}d\eta_{2} \left(\frac{\eta_{1}}{\eta_{2}^4}\right)\bigg{(}\sin^4{\theta} Im \bigg{[}u(\eta_{1})u^{*}(\eta_{e})\bigg{]} Im \bigg{[}u(\eta_{2})u^{*}(\eta_{e})v^{*'}(\eta_{1})v(\eta_{2})\bigg{]}\bigg{)} \nonumber\\
& ~~~~~ = -\frac{31}{245} \e^2 I \epsilon_{H} \sqrt{\frac{6}{\lambda}} \frac{M_{P}}{H} \sin^2{\theta}
\end{align}
\begin{align}
\label{Correction Powerspectrum21}
\frac{\Delta <\overline{\delta \rho}^2(\eta_{e})>_{21}}{<\overline{\delta \rho}^2(\eta_{e})>}
&= \frac{32 \e^2 I \epsilon_{H}\sqrt{6}}{|u^{(0)}(\eta_{e})|^2 \sqrt{\lambda}}\frac{M_{P}}{H}
\int_{\eta_{0}}^{\eta_{e}} d\eta_{1} \int_{\eta_{0}}^{\eta_{1}}d\eta_{2} \left(\frac{\eta_{2}}{\eta_{1}^4}\right)\bigg{(}\sin^4{\theta} Im \bigg{[}u(\eta_{1})u^{*}(\eta_{e})\bigg{]} Im \bigg{[}u(\eta_{2})u^{*}(\eta_{e})v^{*}(\eta_{1})v^{'}(\eta_{2})\bigg{]}\bigg{)} \nonumber\\
& ~~~~~ = -\frac{2}{7} \e^2 I \epsilon_{H} \sqrt{\frac{6}{\lambda}} \frac{M_{P}}{H} \sin^2{\theta}N_{e}
\end{align}
\begin{align}
\label{Correction Powerspectrum22}
\frac{\Delta <\overline{\delta \rho}^2(\eta_{e})>_{22}}{<\overline{\delta \rho}^2(\eta_{e})>}
&= \frac{32 \e^4 I \epsilon_{H}^2}{|u^{(0)}(\eta_{e})|^2 \lambda}\frac{M_{P}^2}{H^2}
\int_{\eta_{0}}^{\eta_{e}} d\eta_{1} \int_{\eta_{0}}^{\eta_{1}}d\eta_{2} \left(\frac{\eta_{e}^4}{\eta_{1}^4\eta_{2}^4}\right)\bigg{(}\sin^4{\theta} Im \bigg{[}u(\eta_{1})u^{*}(\eta_{e})\bigg{]} Im \bigg{[}u(\eta_{2})u^{*}(\eta_{e})v^{*}(\eta_{1})v(\eta_{2})\bigg{]}\bigg{)} \nonumber\\
& ~~~~~ = \frac{9 }{1078} \frac{\e^4 I \epsilon_{H}^2}{\lambda} \frac{M_{P}^2}{H^2} \sin^2{\theta}
\end{align}
where $|u^{(0)}(\eta_{e})|^2 = \frac{1}{2k^3\eta_{e}^2}$ is the amplitude of the free inflaton field fluctuations. Note that, as we showed at the end of the previous Section,  both the inflaton field excitations and the gauge field excitations  remain nearly massless during most of the period of inflation, so we have used the massless mode function approximations for $u_k(\eta)$ and $v_k(\eta)$ given in Eq. (\ref{mode function0}).

The first term, Eq. (\ref{Correction Powerspectrum11}), is the same as in models of real inflaton field  \cite{Watanabe:2010fh}. However, the next three terms Eqs. (\ref{Correction Powerspectrum12}), (\ref{Correction Powerspectrum21}) and
(\ref{Correction Powerspectrum22}) are originated from the interaction $\e^2 \rho^2 A^2$
which does not exist in models with a real inflaton field.  Also note  the relative sign between  Eqs. (\ref{Correction Powerspectrum12}) and  (\ref{Correction Powerspectrum21}) compared to (\ref{Correction Powerspectrum22}).

The contributions of the longitudinal mode, $D_2$ (and $D_2'$) are
\begin{align}
\label{Correction Powerspectrum33}
\frac{\Delta <\overline{\delta \rho}^2(\eta_{e})>_{33}}{<\overline{\delta \rho}^2(\eta_{e})>}
&= \frac{32 \e^4 I \epsilon_{H}^2}{|u^{(0)}(\eta_{e})|^2 \lambda}\frac{M_{P}^2}{H^2}
\int_{\eta_{0}}^{\eta_{e}} d\eta_{1} \int_{\eta_{0}}^{\eta_{1}}d\eta_{2} \left(\frac{\eta_{e}^4}{\eta_{1}^4\eta_{2}^4}\right)\bigg{(}\cos^2{\theta} Im \bigg{[}u(\eta_{1})u^{*}(\eta_{e})\bigg{]} Im \bigg{[}u(\eta_{2})u^{*}(\eta_{e})w^{*}(\eta_{1})w(\eta_{2})\bigg{]}\bigg{)} \nonumber\\
& ~~~~~ = \frac{18}{5} (\e^2 I \epsilon_{H}\lambda) \frac{M_{P}^4}{\cM^4} (k \eta_{e})^2 \cos^2{\theta}
\end{align}

\begin{align}
\label{Correction Powerspectrum34}
\frac{\Delta <\overline{\delta \rho}^2(\eta_{e})>_{34}}{<\overline{\delta \rho}^2(\eta_{e})>}
&= \frac{32 \sqrt{6} \e^2 I \epsilon_{H}}{|u^{(0)}(\eta_{e})|^2 \sqrt{\lambda}}\frac{M_{P}}{H}
\int_{\eta_{c}}^{\eta_{e}} d\eta_{1} \int_{\eta_{c}}^{\eta_{1}}d\eta_{2} \left(\frac{\eta_{2}}{\eta_{1}^4}\right)\bigg{(}\cos^2{\theta} Im \bigg{[}u(\eta_{1})u^{*}(\eta_{e})\bigg{]} Im \bigg{[}u(\eta_{2})u^{*}(\eta_{e})w^{*}(\eta_{1})w^{'}(\eta_{2})\bigg{]}\bigg{)} \nonumber\\
& ~~~~~ = 32 \sqrt{\frac{3\epsilon_{H}}{\lambda}} (\e I \lambda^2) \frac{H M_{P}^4}{\cM^6} (k^2 \eta_{e}) \cos^2{\theta}
\end{align}

\begin{align}
\label{Correction Powerspectrum43}
\frac{\Delta <\overline{\delta \rho}^2(\eta_{e})>_{43}}{<\overline{\delta \rho}^2(\eta_{e})>}
&= \frac{32 \sqrt{6} \e^2 I \epsilon_{H}}{|u^{(0)}(\eta_{e})|^2 \sqrt{\lambda}}\frac{M_{P}}{H}
\int_{\eta_{c}}^{\eta_{e}} d\eta_{1} \int_{\eta_{0}}^{\eta_{1}}d\eta_{2} \left(\frac{\eta_{1}}{\eta_{2}^4}\right)\bigg{(}\cos^2{\theta} Im \bigg{[}u(\eta_{1})u^{*}(\eta_{e})\bigg{]} Im \bigg{[}u(\eta_{2})u^{*}(\eta_{e})w^{'*}(\eta_{1})w(\eta_{2})\bigg{]}\bigg{)} \nonumber\\
& ~~~~~ = -8 \sqrt{\frac{2}{3\lambda}} (I \lambda^2 k^2) \frac{H M_{P}^3}{\cM^4} \left(\frac{3 \e M_{P}\sqrt{2\epsilon_{H}}}{\cM^2} \right)^{2/3} \cos^2{\theta}\left(-\eta_{e}\right)^{4/3}
\end{align}

\begin{align}
\label{Correction Powerspectrum44}
\frac{\Delta <\overline{\delta \rho}^2(\eta_{e})>_{44}}{<\overline{\delta \rho}^2(\eta_{e})>}
&= \frac{192 I}{|u^{(0)}(\eta_{e})|^2} \int_{\eta_{c}}^{\eta_{e}} d\eta_{1} \int_{\eta_{c}}^{\eta_{1}}d\eta_{2} \left(\frac{\eta_{1}\eta_{2}}{\eta_{e}^4}\right)\cos^2{\theta} Im \bigg{[}u(\eta_{1})u^{*}(\eta_e)\bigg{]} Im \bigg{[}u(\eta_{2})u^{*}(\eta_e)w^{*'}(\eta_{1})w^{'}(\eta_{2})\bigg{]} \nonumber\\
& ~~~~~ = \frac{\pi^2}{27}\frac{\e M_{P} \cM\eta_{e}^2}{\Gamma(3/4)^4} \left(I(\frac{3I}{2})^{3/4}\sqrt{\epsilon_{H}}\right) (N_{e} - N_{c})^2 \cos^2{\theta}
\end{align}
Note that the contributions from the longitudinal mode are sourced by $\e$ and are scale-dependent. Furthermore, these terms all have positive powers of $\eta_e$ which are
exponentially small as expected.  As discussed in previous Section there is a cancelation in derivative couplings of the longitudinal mode between the terms coming from the $f^2 F^2$ interaction and a term coming from integrating out $\delta A_0$. As a result,
as shown in  Eq. (\ref{L-ratio}),
the leading interaction from the longitudinal mode during most of the period of inflation ($0< N < N_c$) is much smaller than the leading interaction of the transverse mode.  This justifies why the anisotropy generated from the longitudinal mode is much smaller than the anisotropy generated from the transverse mode.

As a result the  fractional change in the curvature perturbations
power spectrum due to anisotropy, to leading order, is
\ba
\label{partd}
\frac{\Delta <\overline{\delta \rho}^2(\eta_{e})>}{<\overline{\delta \rho}^2(\eta_{e})>} &=& 24 I \sin^2{\theta} N_{e}^2 -\frac{31}{245} \e^2 I \epsilon_{H} \sqrt{\frac{6}{\lambda}} \frac{M_{P}}{H} \sin^2{\theta} -\frac{2}{7} \e^2 I \epsilon_{H} \sqrt{\frac{6}{\lambda}} \frac{M_{P}}{H} \sin^2{\theta}N_{e} \nonumber\\
&& + \frac{9}{1078} \frac{\e^4 I \epsilon_{H}^2}{\lambda} \frac{M_{P}^2}{H^2} \sin^2{\theta} + \frac{18}{5} (\e^2 I \epsilon_{H}\lambda) \frac{M_{P}^4}{\cM^4} (k \eta_{e})^2 \cos^2{\theta} \nonumber\\
&& + 32 \sqrt{\frac{3\epsilon_{H}}{\lambda}} (\e I \lambda^2) \frac{H M_{P}^4}{\cM^6} (k^2 \eta_{e}) \cos^2{\theta} -8 \sqrt{\frac{2}{3\lambda}} (I \lambda^2 k^2) \frac{H M_{P}^3}{\cM^4} \left(\frac{3\e M_{P}\sqrt{2\epsilon_{H}}}{\cM^2} \right)^{2/3} \cos^2{\theta}\left(-\eta_{e}\right)^{4/3} \nonumber\\
&& + \frac{\pi^2}{27}\frac{\e M_{P} \cM\eta_{e}^2}{\Gamma(3/4)^4} \left(I(\frac{3I}{2})^{3/4}\sqrt{\epsilon_{H}}\right) (N_{e} - N_{c})^2 \cos^2{\theta} \, .
\ea
In this formula, $N_e$ stands for the total number of e-folds which we take to be $60$ and $N_c$ is the number of e-fold from the start of inflation till $\eta= \eta_c$ given by Eq. (\ref{Nc}).

As mentioned before, the first four terms in Eq. (\ref{partd}) come from the transverse mode. The first term is similar to
\cite{Watanabe:2010fh} while the next three terms are due to the charge effects which do not exist in models with a real inflaton field.
However, the last four terms in Eq. (\ref{partd}) are due to longitudinal mode which also
do not exist in models with a real inflaton field. However,  since they are suppressed with the powers of $\eta_{e}$ we conclude that their contributions into $g_{*}$ is very small.  As a result, the dominant contribution in $g_*$ comes  from the transverse mode.
Since $\sin^2 \theta = 1- \cos^2 \theta$,
the leading correction to anisotropy power spectrum in  Eq. (\ref{partd})  is
\ba
\label{g-lead}
 g_* \simeq - 24 I N_{e}^2 + \frac{2}{7} \e^2 I \epsilon_{H} \sqrt{\frac{6}{\lambda}} \frac{M_{P}}{H}  N_{e}  - \frac{9}{1078}  \frac{\e^4 I \epsilon_{H}^2}{\lambda} \frac{M_{P}^2}{H^2} \, .
\ea
The interesting thing is that the two contributions of the transverse mode in $g_*$, the last two terms in Eq. (\ref{g-lead}), have different signs. However, one can easily check that  the sign of $g_*$ is always negative, so the positive contribution from the term containing $\e^2$ is always offset by the negative term containing $\e^4$. This is intuitively understandable, since we expect that a total positive contribution in $g_*$ comes from the longitudinal mode which are exponentially suppressed in this model while we do not expect the net contribution from the transverse mode
to give a positive contribution in $g_*$. This is consistent with the results in \cite{Watanabe:2010fh}.

Demanding that $|g_*| < 0.3$ in order not to produce too much anisotropy,
we find that  $I\simeq 10^{-5}$ and \textbf{$ \e^2 \leq 10 \sqrt{\frac{\lambda}{I \epsilon_{H}^2}} \frac{H}{M_{P}}$}.
For typical model parameters in symmetry breaking inflation, this leads to $\e \lesssim 10^{-3}$.

As observed in \cite{Bartolo:2012sd}  the infra-red (IR) modes of the vector field perturbations remain frozen on super-horizon scales which accumulate to renormalize the background gauge field. As a result, this can lead to a large value of $g_*$ unless one takes
$N_e \sim 60$ as we have assumed here.

 %%%%%%%%%%%%%%%%%%%%%%%%%%%%%%%%%%%%%%%%%%%%%%%%
\section{The Origin of the Leading Interactions Terms}
\label{matter-leading}

Having calculated the anisotropic power spectrum through complicated procedure of integrating out
the non-dynamical fields and approximating $\bar \lambda_1$ and other $\lambda_i$ and
$\bar \lambda_i$, one may wonder what the origins of the leading interaction terms $L_{\rho D_1}$ and $L_{\rho D_2}$, or alternatively $L_{\rho A_{1} }$, $L_{\rho M}$  and $L_{A_{1} M}$, are. Are they coming from the metric perturbations or from the matter sector? 

The full second order action containing both the matter perturbations  and the metric perturbations contributions are given in Appendix \ref{reduced}. Subsequently, in Appendix \ref{second slow} we have presented the leading order actions in slow-roll approximation which were used in Section \ref{eta-less} and \ref{eta-big}. Here we show that these leading interactions actually come from the matter perturbations. In other words, below we show that the contributions of the  matter sector are actually the same leading terms which were used in
Section \ref{eta-less} and \ref{eta-big}.

To show this first we integrate out $\delta A_0$ and then read off the interaction terms containing the matter perturbations.  The leading terms in the matter sector coming from integrating out $\delta A_{0}$ are
%\textit{\textbf{Leading terms from $\delta A_{0}$:}}
\begin{align}
\label{lead A0}
&-\frac{k^2 \cos^2{\theta}}{\bar \lambda_{1}}\frac{b^4}{2a^3}\frac{\sqrt{6I}}{\eta}f^3(\delta \rho^* \delta A_{1}' + c.c.)-\frac{k^3\cos{\theta} \sin^2{\theta}} {\bar \lambda_{1}}\frac{b^4}{2a^3}\frac{\sqrt{6I}}{\eta}f^3(i\delta \rho^* M' + c.c.)\nonumber\\
&-\frac{b^4}{4a^4}\frac{k^3}{\bar \lambda_{1}} f^4 \sin^2{\theta}\cos{\theta}(i M' \delta A_{1}^{'*} + c.c.) \, .
\end{align}
On the other hand, the leading terms for the matter perturbations present in the original
action  (without  integrating out any fields) are
%\textit{\textbf{Leading terms from action:}}
\begin{align}
\label{lead action}
\frac{b^2}{a}\frac{\sqrt{6I}}{\eta}f (\delta \rho^* \delta A_{1}' + c.c.)+ \frac{b^2}{2a^2}k^3 f^2 \sin^2{\theta}\cos{\theta}(iM \delta A_{1}^{*} + c.c.)
-\e^2b^2 \rho A_{x}\Big{(}\delta \rho \delta A_{1}^{*} + c.c. \Big{)}\, .
\end{align}
So by adding Eq. (\ref{lead action}) and (\ref{lead A0}) we can obtain all the leading interaction terms  for $L_{\rho A_{1} }$ , $L_{\rho M}$ and $L_{A_{1} M}$ as,
\begin{align}
\label{Leading Terms}
&\left(\frac{b^2}{a}\frac{\sqrt{6I}}{\eta}f-\frac{k^2 \cos^2{\theta}}{\bar \lambda_{1}}\frac{b^4}{2a^3}\frac{\sqrt{6I}}{\eta}f^3\right)\bigg{(}\delta \rho^* \delta A_{1}' + c.c. \bigg{)}-\frac{k^3\cos{\theta} \sin^2{\theta}} {\bar \lambda_{1}}\frac{b^4}{2a^3}\frac{\sqrt{6I}}{\eta}f^3\bigg{(}i\delta \rho^* M' + c.c.\bigg{)}\nonumber\\
&+ \left(\frac{b^2}{2a^2}k^3 f^2 \sin^2{\theta}\cos{\theta}\right)\bigg{(}iM \delta A_{1}^{*} + c.c.\bigg{)}-\left(\frac{b^4}{4a^4}\frac{k^3}{\bar \lambda_{1}} f^4 \sin^2{\theta}\cos{\theta}\right)\bigg{(}iM' \delta A_{1}^{*'} + c.c.\bigg{)}-\e^2b^2 \rho A_{x}\Big{(}\delta \rho \delta A_{1}^{*} + c.c. \Big{)}\, .
\end{align}
Interestingly, this is the whole leading action which was used in previous sections
to calculate the anisotropic power spectrum.

As a result, the leading interaction terms for the first phase, $\eta < \eta_c$, are
\ba
\label{Leading Terms1}
L_{\mathrm{lead.}}&=& \left(\frac{1}{\eta}\right)\left(\frac{b^2f}{a}\sqrt{6I}\sin^2{\theta}\right)\left(\delta \rho ^{*} D_{1}'+c.c. \right) - \left( \frac{a^2}{f\eta} \right) \e^2 \sqrt{\frac{I \epsilon_{H}^2}{\lambda}}M_{P} \sin^2{\theta}\Big{(} \delta \rho^{*} D_{1} + c.c.\Big{)}\nonumber\\
&& - \left( \frac{a^2}{f\eta} \right) \e^2 \sqrt{\frac{I \epsilon_{H}^2}{\lambda}}M_{P} \cos{\theta}\Big{(} \delta \rho^{*} D_{2} + c.c.\Big{)}
\ea
Interestingly, this is exactly the leading term interaction as obtained in Eq. (\ref{rhoD1new}). Similarly, for the second phase, $\eta > \eta_c$, Eq. (\ref{Leading Terms}) yields
\ba
\label{Leading Terms2}
L_{\mathrm{lead.}}&=& \left(\frac{1}{\eta}\right)\left(\frac{b^2f}{a}\sqrt{6I}\sin^2{\theta}\right)\left(\delta \rho ^{*} D_{1}'+c.c. \right) - \left( \frac{a^2}{f\eta} \right) \e^2 \sqrt{\frac{I \epsilon_{H}^2}{\lambda}}M_{P} \sin^2{\theta}\Big{(} \delta \rho^{*} D_{1} + c.c.\Big{)}\nonumber\\
&& -  \left( \frac{a^2}{f\eta} \right) \e^2 \sqrt{\frac{I \epsilon_{H}^2}{\lambda}}M_{P} \cos{\theta}\Big{(} \delta \rho^{*} D_{2} + c.c.\Big{)}
+\left(\frac{1}{\eta}\right)\left(\frac{b^2f}{a}\sqrt{6I}\cos{\theta}\right)\left(\delta \rho ^{*} D_{2}'+c.c. \right) \, .
\ea
As expected, this expression is the sum of the leading interaction terms Eqs. (\ref{rhoD1 action20}) and (\ref{rhoD2 action20}).

In summary we conclude that the leading interactions in generating anisotropies originate from the matter sector and one can neglect the metric perturbations in calculating the leading order corrections to the curvature perturbations power spectrum . Computationally, this is a very important result which considerably simplifies the perturbation analysis in similar models. This conclusion was also reached in  \cite{Bartolo:2012sd}.

This also explains why in the processes of integrating out the non-dynamical fields only  $\bar \lambda_1$ plays prominent roles. As mentioned below Eq. (\ref{b-lambda1}) $\bar \lambda_1$ originates from integrating out $\delta A_0$ which is the non-dynamical field in the matter sector. On the other hand,
other $\lambda_i$ and $\bar \lambda_i$ originate from integrating out the non-dynamical fields
$A, B$ and $\beta$ in the metric side which should not play prominent roles as expected from the above results.

%%%%%%%%%%%%%%%%%%%%%%%%%%%%%%%%%%%%%%%%%%%%%%%%%%
\section{Summary and Discussions}
\label{Summary}

In this work we have studied anisotropy generated in  an anisotropic inflationary scenario
with a complex scalar field charged under the $U(1)$ gauge field. Because of the Abelian Higgs mechanism, the gauge field obtains the dynamical mass $\e^2 \rho^2 A_\mu A^\mu$. As a result,  the angular excitations of the complex scalar field is eaten by the gauge field so the longitudinal component of $A_\mu$ becomes excited.

There are two types of interactions in the system. The first interaction originates from the gauge kinetic coupling $f(\rho)^2 F^2$ while the second interaction comes from the symmetry breaking effect $\e^2 \rho^2 A_\mu A^\mu$. These interactions induce exchange vertices  between $\delta \rho$ and the transverse and the longitudinal modes encoded in  the
interactions $L_{\rho D_1}$ and $L_{\rho D_2}$.   As discussed in details in Section \ref{sec-ac} the dominant interaction during the period $0< N< N_c$ is $L_{\rho D_1}$ originated from
$f(\rho)^2 F^2$ which is similar to models with a real inflaton field. As a result the leading exchange vertex is given by the derivative coupling of the transverse mode. However, during the phase $N_c \leq N \leq N_e$ the dominant interaction is given by $\e^2 \rho^2 A_\mu A^\mu$. Correspondingly, the dominant exchange vertices are the terms in $L_{\rho D_1}$ and $L_{\rho D_2}$ containing the coupling $\e^2$.

The leading contributions to anisotropic power spectrum are  given in Eq. (\ref{partd}) and
Eq.  (\ref{g-lead}). The first four terms in  Eq. (\ref{partd}) come from the interaction of $\delta \rho $ with the transverse mode, $L_{\rho D_1}$.
In terms of Feynman diagrams this interaction is represented  by the exchange vertex shown in
Fig. 1 (a).   This is similar to the result obtained  in  \cite{Watanabe:2010fh} plus the contributions in Eq.  (\ref{g-lead}) containing the effects of $\e$. As we showed, the sign of $g_*$ is always negative.  In order to satisfy the observational constraints on curvature perturbation power spectrum we obtain $I \lesssim 10^{-5}$ and $\e \lesssim 10^{-3}$.
In addition, unlike \cite{Watanabe:2010fh}, the longitudinal mode $D_2$ also contributes into the anisotropic power spectrum.  In terms of the Feynman diagrams this interaction is represented  by the exchange vertex shown in Fig. 1 (b). However,  the longitudinal mode contributes only towards the end of inflation  and  its contributions to the anisotropic power spectrum are hugely suppressed  compared to the contribution from the transverse mode. 

We also verified that the leading interactions in anisotropic power spectrum come from the matter sector perturbations. In other words, to calculate the leading order corrections into the power spectrum, one can neglect the metric perturbations. Computationally, this knowledge simplifies the analysis considerably. This is particularly helpful when calculating the bispectrum and non-Gaussianities which we would like to come back in a future work.

The issue of generating statistical anisotropy at the end of inflation via waterfall dynamics have been considered in \cite{Yokoyama:2008xw}. However, it is shown in \cite{Emami:2011yi} that  this mechanism does not work and the anisotropy produced purely from waterfall effect
at the end of inflation  is exponentially suppressed.  This conclusion, however, was criticized in \cite{Lyth:2012vn}. Having this said, we believe that the conclusion derived in \cite{Emami:2011yi},  which is obtained by a careful use of $\delta N$ formalism, is valid. To get large enough statistical anisotropy in model of \cite{Yokoyama:2008xw}, one has to consider the evolution of the gauge field both at the background level and at the perturbation level during the entire inflationary period, as we did here.  This point was also mentioned in \cite{Bartolo:2012sd}. We would like to pursue  this issue in a future work considering the charged hybrid model using the standard in-in formalism as employed here.

In this work we have only calculated anisotropy in  curvature perturbation power spectrum. However, after inflation ends the Universe becomes isotropic. As a result, we restore the usual two degrees of freedom associated with the tensor perturbations. One can specifically check that the scalar perturbation $\gamma$ and the vector perturbations $\Gamma_i$ furnish two polarizations of tensor perturbations after inflation ends. Note that $\Gamma_i$, subject to 
$\partial _i \Gamma_i =0$ during anisotropic inflation, has only one degrees of freedom 
($\Gamma_3$ in our convention) so it can account only for one tensor polarization while the other polarization is given by $\gamma$ as expected. As shown in Appendices \ref{reduced}
and \ref{second slow} the interactions  $L_{\delta \rho \gamma}, L_{\gamma D_1 }$ and $L_{\gamma D_2}$ are generated in our system. As a result  there will be cross correlation between the tensor and scalar perturbations in the form of $\langle \delta \rho \gamma \rangle $ as studied in \cite{Watanabe:2010fh}. Following the in-in formalism analysis,  the cross correlation $\langle \delta \rho \gamma \rangle $ has contributions from the interaction  $L_{\delta \rho \gamma}$ and also contributions  from the second order action
$ L_{\delta \rho D_1} L_{\gamma D_1 } $. As in \cite{Watanabe:2010fh} we expect to have a contribution  like $- 24 I \sqrt{\epsilon_H} N^2 \sin^2 \theta $ in $ \langle \delta \rho \gamma \rangle$.  In addition, our analysis shows that we also obtain contributions proportional to $\e^2$ and $\e^4$ with the structure similar to the corresponding 
terms in $g_*$ in Eq. (\ref{g-lead}).  A complete analysis of the scalar and tensor perturbations cross-correlation  is an interesting question which is beyond the scope of this work. We would like to come back to this question in a future work.

%%%%%%%%%%%%%%%%%%%%%%%%%%%%%%%%%%%%%%%%%%%%%%%%%
\acknowledgements{We would like to thank X. Chen,  K. Dimopoulos, A. Ricciardone and  J. Soda for useful discussions. We  thank M. Peloso for useful comments on the draft and for helpful discussions. We also thank the anonymous referee for the careful comments on the draft and for the insightful hints on the importance of the tensor and scalar cross-correlations.  H.F. would like to thank  the Yukawa Institute for Theoretical Physics at Kyoto University for the hospitality where this work was in progress during the Long-term Workshop YITP-T-12-03
on ``Gravity and Cosmology 2012''. R.E. is very gratefull to ICTP for their warm hospitality when the corrections of this work was done. }

%%%%%%%%%%%%%%%%%%%%%%%%%%%%%%%%%%%%%%%%%%%%%%%%%%
\appendix

\section{Metric Perturbations}
\label{appendix1}

Here we study the metric perturbations in Bianchi I background and their transformation properties  under a general coordinate transformation. Consider the general coordinate transformation
\ba
\label{xi}
x^\mu \rightarrow x^{\mu} + \xi^{\mu} \quad \quad ,  \quad \quad
\xi^\mu = \left(  \xi^0 \, ,\,  \partial_i \lambda  \, ,\, \partial_i \Lambda + \xi_\perp^i
\right)
\ea
in which $\xi^0, \lambda$ and $\Lambda$ are scalars and $\xi_\perp^i$ is vector subject to
$\partial_i {\xi_\perp^i}$=0.  For the future reference note that by appropriate choice of $\xi^0, \lambda$ and $\Lambda$   one can remove three scalar degrees of metric perturbations in Eq. (\ref{deltag}) while the freedom from $\xi_\perp^i$ can remove only one  vector degree of freedom.

Under the coordinate transformation Eq. (\ref{xi}) we have
\ba
\delta g_{\mu \nu} \rightarrow \delta g_{\mu \nu} -{^{(0)}} g_{ \mu \nu, \kappa}\, \xi^{\kappa} -
{^{(0)}}g_{\alpha \nu}\,  \partial_\mu \xi^\alpha -  {^{(0)}}g_{\alpha \mu}\,  \partial_\nu \xi^\alpha
\ea
in which ${^{(0)}}g_{\alpha \mu}$ is the background Bianchi metric given in Eq. (\ref{Bianchi-metric}).

More explicitly, one can check that
\ba
A && \rightarrow A - \frac{1}{a} \left( a \xi^0 \right)'   \\
\beta&& \rightarrow \beta + \xi^0 - \lambda'\\
B &&\rightarrow  B+ \frac{a}{b} \xi^0 - \frac{b}{a} \Lambda'\\
\bar \psi && \rightarrow \bar \psi +  \frac{a'}{a} \xi^0 + \partial_x^2 \lambda\\
\gamma && \rightarrow \gamma - \frac{b}{a}  \Lambda - \frac{a}{b}   \lambda\\
\psi  && \rightarrow \psi  + \frac{b'}{b} \xi^0 \\
E && \rightarrow E - \Lambda
\ea
and
\ba
B_i && \rightarrow B_i - \frac{b}{a}\,  {\xi_\perp^{i'}} \\
\Gamma_i && \rightarrow  \Gamma_i  - \frac{b}{a}\,  {\xi_\perp^i} \\
E_i && \rightarrow E_i - \xi_\perp^i
\ea
Using the above transformation properties one can check that the following two scalar variables are gauge invariant
\ba
\label{delta-phi-psi}
\delta \phi_\psi &&\equiv \delta \phi + \frac{\dot \phi}{H_b} \psi  \\
\hat \gamma &&\equiv \gamma - \frac{b}{a} E + \frac{a}{b} \partial^{-2}_{1} \left( \bar \psi - \frac{H_a}{H_b} \psi
\right)
\ea
In this view $\delta \phi_\psi$ represents the inflaton perturbations on  $\psi=0$ surface which
reduces to  inflaton perturbations on flat slice in FRW background while $\hat \gamma$ is identically zero in FRW background.

In our analysis we adopt the following gauge
\ba
\label{gauge}
\psi= \bar \psi = E = E_i =0 \, ,
\ea
which one can check is a consistent gauge.  Note that the three scalar conditions $\psi= \bar \psi = E =0$ fixes three scalar freedoms $\xi^0, \lambda $ and $\Lambda$ while the vector condition $E_i=0$ fixes the remaining one degree of freedom $\xi_\perp$.
The advantage in choosing the gauge in Eq. (\ref{gauge}) is that it reduces to the flat gauge in
the isotropic limit where $\psi =\bar \psi$.

We also note that after inflation ends and the universe becomes isotropic the scalar perturbation $\gamma$ and the vector perturbation $\Gamma_i$ combine to furnish two polarizations of the tensor perturbations. Note that in anisotropic background, the condition
$\partial_i \Gamma_i=0$ leaves only one degree of freedom. As a result $\Gamma_i$ can count only for one tensor polarization and the remaining polarization is taken care of by
$\gamma$ as mentioned.

%%%%%%%%%%%%%%%%%%%%%%%%%%%%%%%%%%%%%%%%%%%%%%%%%%
\section{Integrating out non-Dynamical Fields}
\label{reduced}

In this appendix we present the detail analysis of integrating out the non-dynamical fields $\delta A_0, \beta, A$ and $B$ in terms of the dynamical fields $\delta \rho, \gamma, \delta A_1$ and $M$. The second order action is given in Eq. (\ref{S2-scalar}).  Correspondingly, the second order action for the scalar perturbations in Fourier space is

\ba
\label{S2-scalar-k}
&&S_2 = \int d \eta d^3 k \left[ b b' k_x^2 (A^* \beta + A \beta^*) + \frac{a b}{2} (\frac{a'}{a} + \frac{b'}{b}) k_y^2 (A^* B + A B^*) +  \frac{a b}{2} k_x^2 k_y^2 (\gamma^* A + \gamma A^*) - a^2 b^2 V(\rho_0) |A|^2
 \right. \nonumber \\ &&\left.
- \frac{\e^2}{2} b^2 \rho_0^2 A_x^2 |A|^2 - \frac{a b}{4} k_x^2 k_y^2 (\beta^* B + \beta B^*) + \frac{a' b}{2} k_x^2 k_y^2 (\gamma^* \beta + \gamma \beta^*) + \frac{a b}{4} k_x^2 k_y^2 (\gamma^* \beta' + \gamma \beta'^*)
+ \frac{\e^2}{2} b^2 \rho_0^2 A_x^2 k_x^2 |\beta|^2
\right. \nonumber \\ &&\left.
+ \frac{a^2}{4} k_x^2 k_y^2 |\beta|^2 - \frac{b^2}{4} k_x^2 k_y^2 (B^* \gamma' + B \gamma'^*) + \frac{b^2}{4} (\frac{b'}{b} - \frac{a'}{a}) k_x^2 k_y^2 (\gamma^* B + \gamma B^*) + \frac{b^2}{4} k_x^2 k_y^2 |B|^2 + \frac{b^2}{4} k_x^2 k_y^2 | \gamma'|^2
\right. \nonumber \\ &&\left.
- \frac{\e^2}{2}b^2 A_x^2 \rho_0^2 k_x^2 k_y^2 |\gamma|^2 + \frac{f^2 b^2}{2 a^2} A_x'^2 k_x^2 k_y^2 |\gamma|^2 - \frac{b^2}{4} (\frac{b''}{b} - \frac{a''}{a}) k_x^2 k_y^2 |\gamma|^2
+  \frac{b^2}{2} |\delta \rho'|^2 -  \frac{b^2}{2} \rho_0' (A^* \delta \rho' + A \delta \rho'^*)
\right. \nonumber \\ &&\left.
- \frac{b^2}{2} \rho_0' k_x^2 (\beta^* \delta \rho + \beta \delta \rho^*) - \frac{a b}{2} \rho_0' k_y^2 (B^* \delta \rho + B \delta \rho^*) - \frac{b^2}{2} k_x^2 | \delta \rho|^2 - \frac{a^2}{2} k_y^2 |\delta \rho|^2 + \frac{\e^2 b^2}{2}  \rho_0^2 |\delta A_0|^2
\right. \nonumber \\ &&\left.
+ i k_x \frac{\e^2 b^2}{2} \rho_0^2 A_x  (\beta^* \delta A_0 - \beta \delta A_0^*)
-\frac{\e^2 b^2}{2} \rho_0^2 |\delta A_1|^2 - \frac{\e^2 b^2}{2}A_x^2 |\delta \rho|^2
- \e^2 b^2 \rho_0 A_x (\delta \rho^* \delta A_1 + \delta \rho \delta A_1^*)
\right. \nonumber \\ &&\left.
+ i k_x k_y^2 \frac{\e^2 a b}{2} \rho_0^2 A_x (\gamma M^* - \gamma^* M) - \frac{\e^2 a^2}{2} \rho_0^2 k_y^2 |M|^2 - \frac{\e^2 b^2}{2} \rho_0^2 A_x (A^* \delta A_1 + A \delta A_1^*)
- \frac{\e^2  b^2}{2} \rho_0 A_x^2 (A^* \delta \rho + A \delta \rho^*)
\right. \nonumber \\ &&\left.
+ \frac{b^2}{2 a^2} f^2  |\delta A_1'|^2 +   \frac{b^2}{2 a^2} f^2  k_x^2 |\delta A_0|^2
- i k_x \frac{b^2 f^2}{2 a^2}  (\delta A_1'^* \delta A_0 - \delta A_1' \delta A_0^*)
-  \frac{b^2 f^2}{2 a^2} A_x' (A^* \delta A_1' + A \delta A_1'^*)
\right. \nonumber \\ &&\left.
+i k_x  \frac{b^2 f^2}{2 a^2} A_x' (A^* \delta A_0 - A \delta A_0^*) - i k_x k_y^2 \frac{b f^2 A_x'}{2a}  (\gamma M'^* - \gamma^* M')   + i k_x k_y^2 \frac{b f^2}{2 a} A_x'   (\gamma \delta A_0^* - \gamma^* \delta A_0)
\right. \nonumber \\ &&\left.
-\frac{b f^2}{2 a} A_x' k_y^2 (B \delta A_1^* + B^* \delta A_1)  +  i k_x\frac{b f^2}{2 a} A_x' k_y^2 (B^* M - B M^*) + \frac{f^2}{2} k_y^2 |M'|^2 + \frac{f^2}{2} k_y^2| \delta A_0 |^2
\right. \nonumber \\ &&\left.
- \frac{f^2}{2} k_y^2 (M'^* \delta A_0 + M' \delta A_0^*) - \frac{f^2}{2} k_y^2 |\delta A_1 |^2
- \frac{f^2}{2} k_x^2 k_y^2 |M |^2 + i k_x \frac{f^2}{2} k_y^2 (\delta A_1^* M - \delta A_1 M^*)
\right. \nonumber \\ &&\left.
+ \frac{b^2 f f_{, \rho}}{a^2}  A_x' (\delta A_1'^* \delta \rho + \delta A_1' \delta \rho^*)
 + i k_x \frac{b^2 f f_{, \rho}}{a^2}  A_x' (\delta A_0^* \delta \rho - \delta A_0 \delta \rho^*) - \frac{b^2 f f_{, \rho}}{2 a^2} A_x'^2 ( A^* \delta \rho +  A \delta \rho^*)
\right. \nonumber \\ &&\left.
+ \frac{b^2  f_{, \rho}^2}{2 a^2} A_x'^2 |\delta \rho |^2 +  \frac{b^2 f f_{, \rho \rho}}{2 a^2} A_x'^2 |\delta \rho |^2 - \frac{a^2 b^2}{2} V_{, \rho \rho} |\delta \rho |^2  -  \frac{a^2 b^2}{2} V_{, \rho } (\delta \rho A^* + \delta \rho^* A)
\right] \, .
\ea

We have to integrate out the non-dynamical variables $\{\delta A_0, \beta, A, B\}$ from the action Eq. (\ref{S2-scalar-k}). The analysis are simple but tedious. To outline the analysis, here we demonstrate how to integrate out $\beta$. The action expanded in powers of $\beta$ is
\ba
\label{L-beta}
{\cal L } = c_1 \beta \beta^* + c_2 \beta^* + c_2^* \beta  +... \, ,
\ea
in which the dots indicates the rest of the action containing the dynamical fields $\{\delta \rho, \delta A_1, M, \gamma \}$ and $\{ \delta A_0, A, B\}$ and $c_{1, 2}$ are functions which can be read off from the action Eq. (\ref{S2-scalar-k})
\ba
c_1 = \frac{a^2}{4} k_x^2 k_y^2 + \frac{\e^2}{2} b^2 \rho^2 k_x^2 A_x^2
\ea
and
\ba
c_2 = b \, b' k_x^2 A - \frac{ab}{4} k_x^2 k_y^2 \, B + \frac{a' b}{2} k_x^2 k_y^2 \, \gamma
- k_x^2 k_y^2 \left( \frac{ab}{4} \right)' - \frac{b^2}{2} \rho' k_x^2 \delta \rho
+ i k_x \frac{\e^2 b^2}{2} \rho^2 A_x \delta A_0 \, .
\ea
Varying the action with respect to $\beta^*$ yields $\beta = - c_2/c_1$. Plugging this into the action yields
\ba
{\cal L} = -\frac{|c_2|^2}{c_1} + ... .
\ea
Following the same steps to integrate out $\delta A_0, A$ and $B$ we can write the dynamical action as
\ba
L_{(2)} =  L_{\rho \rho} + L_{\gamma \gamma} + L_{MM}  + L_{A_1 A_1} + L_{\rho \gamma} + L_{\rho M} + L_{\rho A_1} + L_{\gamma M} + L_{\gamma A_1}
+ L_{M A_1}  + ... \, ,
\ea
in which the dots indicate the rest of the action coming from the dynamical fields $\{\delta \rho, \delta A_1, M, \gamma \}$. Here we have defined

%\begin{align}
\ba
\label{L_{rho-rho}}
L_{\rho \rho}  &&= \left(\frac{b^2}{2} - \frac{1}{\lambda_8}\frac{b^4}{4}\rho_{0}^{'2}- \frac{\lambda_6^2}{\lambda_{13} \lambda_8^2}\frac{b^4}{4}\rho_{0}^{'2}  \right) \Big| \delta \rho ^{'}\Big|^2+ \bigg{(}-\frac{b^2}{2}k^2 - \frac{b^2}{2}\e^2 A_{x}^2+ \frac{b^2}{2a^2}A_{x}^{'2}f_{,\rho}^2
 + \frac{b^2}{2a^2}A_{x}^{'2}f f_{,\rho\rho} - \frac{a^2 b^2}{2}V_{, \rho \rho} \nonumber \\
&&- \frac{1}{\lambda_1}\frac{b^4}{4}\rho_{0}^{'2}k_{x}^4 - \frac{|\lambda_5|^2}{\lambda_2}- \frac{|\lambda_9|^2}{\lambda_8}- \frac{|\lambda_{14}|^2}{\lambda_{13}}\bigg{)}\Big| \delta \rho \Big|^2 + %\nonumber\\
 \bigg{(}  \frac{\lambda_9}{\lambda_8}\frac{b^2}{2}\rho_{0}^{'} - \frac{\lambda_6 \lambda_{14}}{\lambda_8 \lambda_{13}}\frac{b^2}{2}\rho_{0}^{'}\bigg{)}\Big( \delta \rho \delta \rho^{*} \Big)^{'}
\label{L_{gamma-gamma}}
\ea
\ba
L_{\gamma \gamma} & =& \left(\frac{b^2}{4}k_{x}^2 k_{y}^2 -\frac{a^2 b^2}{16}\frac{1}{\bar \lambda_2}k_{x}^4 k_{y}^4- \frac{a^2 b^2}{16}\frac{\bar \lambda_3^2}{\lambda_{8}\bar \lambda_2^2 }k_{x}^4 k_{y}^4 - \frac{|\lambda_{11}|^2}{\lambda_{13}} \right) \Big| \gamma ^{'}\Big|^2+ \bigg{(}-\frac{b^2}{2}\e^2 A_{x}^2 \rho_{0}^2 k_{x}^2 k_{y}^2 + \frac{b^2}{2a^2}A_{x}^{'2}f^2 k_{x}^2 k_{y}^2 \nonumber \\
& - &\frac{b^2}{4}(\frac{b^{''}}{b}-\frac{a^{''}}{a})k_{x}^2 k_{y}^2 - \frac{b^2}{4a^2}\frac{1}{\bar \lambda_1}f^4 A_{x}^{'2} k_{x}^2 k_{y}^4 - \frac{(\bar \lambda_4)^2}{\bar \lambda_2} - \frac{\lambda_7^2}{\lambda_8} -\frac{\lambda_{12}^2}{\lambda_{13}} \bigg{)}\Big| \gamma \Big|^2 +
\bigg{(} \frac{a b}{4}\frac{\bar \lambda_4}{\bar \lambda_2}k_{x}^2 k_{y}^2 - \frac{a b}{4}\frac{\lambda_7 \bar \lambda_{3}}{\lambda_8 \bar \lambda_{2}}k_{x}^2 k_{y}^2 \nonumber \\
&- &\frac{\lambda_{11} \lambda_{12}}{\lambda_{13} }\bigg{)}\Big( \gamma \gamma^{*} \Big)^{'}
%\end{align}
\ea
\ba
\label{L_{M-M}}
L_{MM} & =& \left(\frac{f^2}{2}k_{y}^2 -\frac{f^4}{4}\frac{1}{\lambda_2}k_{y}^4- \frac{f^4}{4}\frac{ |\lambda_3|^2}{\lambda_{8}\lambda_2^2 }k_{y}^4 - \frac{|\lambda_{17}|^2}{\lambda_{13}} \right) \Big| M ^{'}\Big|^2+ \bigg{(}-\frac{a^2}{2}\e^2\rho^2 k_{y}^2 - \frac{f^2}{2}k_{x}^2 k_{y}^2
 - \frac{b^2}{4a^2\lambda_{13}}f^4 A_{x}^{'2} k_{x}^2 k_{y}^4 \bigg{)}\Big| M \Big|^2 \nonumber \\
&& + \bigg{(} \frac{b}{2a\lambda_{13}}f^2 A_{x}^{'}k_{x} k_{y}^2\bigg{)}\Big(i\lambda_{17} M^{*} M^{'}- i\lambda_{17}^{*} M^{'*} M \Big)
\ea
\ba
\label{L_{A1-A1}}
L_{A_{1}A_{1}}& = &\left(\frac{b^2}{2a^2}f^2-\frac{b^4}{4a^4\lambda_2}f^4k_{x}^2- \frac{ (\lambda_{10})^2}{\lambda_{8}} - \frac{(\lambda_{16})^2}{\lambda_{13}} \right) \Big| \delta A_{1} ^{'}\Big|^2+ \bigg{(}-\frac{b^2}{2}\e^2\rho^2 - \frac{f^2}{2}k_{y}^2 - \frac{b^4}{4\lambda_{8}}e^4 \rho^4 A_{x}^{2}
 - \frac{(\lambda_{15})^2}{\lambda_{13}} \bigg{)}\Big| \delta A_{1} \Big|^2 \nonumber \\
 &&+\bigg{(} \frac{b^2\lambda_{10}}{2\lambda_{8}}\e^2 \rho^{2} A_{x} - \frac{\lambda_{15}\lambda_{16}} {\lambda_{13}}\bigg{)}\Big( \delta A_{1} \delta A_{1}^{*} \Big)^{'}
\ea
\ba
\label{L_{rho-gamma}}
L_{\rho \gamma }& =& \Big{(}\frac{b^3}{2a^3\bar \lambda_{1}}f^3 f_{,\rho}A_{x}^{'2}k_{x}^2 k_{y}^2 - \frac{\bar \lambda_4 \bar \lambda_5}{\bar \lambda_2}- \frac{\lambda_7 \lambda_9}{\lambda_8}-\frac{\lambda_{12} \lambda_{14}}{\lambda_{13}}\Big{)} \Big{(}\delta \rho \gamma^{*} + c.c. \Big{)}+ \Big{(} \frac{ab}{4}\frac{\bar \lambda_{5}}{\bar \lambda_{2}}k_{x}^2 k_{y}^2 - \frac{ab}{4}\frac{\bar \lambda_{3}\lambda_{9}}{\bar \lambda_{2}\lambda_{8}}k_{x}^2 k_{y}^2 \nonumber \\
-&& \frac{\lambda_{11}\lambda_{14}}{\lambda_{13}} \Big{)}\Big{(} \delta \rho \gamma^{'*}+c.c. \Big{)}
+ \Big{(} \frac{b^2}{2}\frac{\lambda_{7}}{\lambda_{8}}\rho^{'} - \frac{b^2}{2}\frac{\lambda_{6}\lambda_{12}}{\lambda_{8}\lambda_{13}} \rho^{'}\Big{)}\Big{(}\delta \rho^{'} \gamma^{*} + c.c. \Big{)}+ \Big{(} \frac{ab^3}{8}\frac{\bar \lambda_3}{\bar \lambda_2 \lambda_8} \rho^{'} k_{x}^2 k_{y}^2\nonumber\\
&-& \frac{b^2}{2}\frac{\lambda_6 \lambda_{11}}{\lambda_8 \lambda_{13}} \rho^{'}\Big{)}
\Big{(}\delta \rho^{'} \gamma^{'*} + c.c. \Big{)}
\ea
\ba
\label{L_{rho-M}}
L_{\rho M }& =& \Big{(}\frac{b}{2a}\frac{\lambda_{14}}{\lambda_{13}}f^2 A_{x}^{'}k_{x}k_{y}^2\Big{)} \Big{(}i\delta \rho M^{*} + c.c. \Big{)}+ \Big{(} \frac{f^2}{2\lambda_{2}}k_{y}^2\Big{)}\Big{(}\lambda_{5}\delta \rho M^{'*} + c.c. \Big{)}-\Big{(}\frac{\lambda_{9}}{2\lambda_{8}\lambda_{2}} f^2 k_{y}^2\Big{)}\Big{(}\lambda_{3}\delta \rho M^{'*}\nonumber\\
&+& c.c. \Big{)}-\Big{(}\frac{\lambda_{14}}{\lambda_{13}}\Big{)}\Big{(}\lambda_{17}\delta \rho^{*} M^{'}+ c.c. \Big{)}+ \Big{(}\frac{b^3}{4a}\frac{\lambda_{6}}{\lambda_{13}\lambda_{8}}f^2 A_{x}^{'}\rho^{'}k_{x}k_{y}^2\Big{)} \Big{(}i\delta \rho^{'} M^{*} + c.c. \Big{)}+\Big{(}\frac{b^2}{4\lambda_{8}\lambda_{2}} f^2 k_{y}^2\rho^{'} \Big{)}\nonumber\\
&&\Big{(}\lambda_{3}\delta \rho^{'} M^{'*}+ c.c.\Big{)}- \Big{(}\frac{b^2\lambda_{6}}{2\lambda_{8}\lambda_{13}}\rho^{'} \Big{)}\Big{(}\lambda_{17}\delta \rho^{'*} M^{'}+ c.c.\Big{)}
\ea
\ba
\label{L_{rho-A1}}
L_{\rho A_{1} }& = &\Big{(}-\e^2b^2 \rho A_{x} + \frac{b^2\lambda_{9}}{2\lambda_{8}}\e^2 \rho^2 A_{x} - \frac{\lambda_{14}\lambda_{15}}{\lambda_{13}}\Big{)} \Big{(}\delta \rho \delta A_{1}^{*} + c.c. \Big{)}+ \Big{(} \frac{b^2}{a^2}A_{x}^{'}f f_{,\rho}- \frac{b^2}{2a^2}\frac{i\lambda_{5}^{*}}{\lambda_{2}}k_{x}f^2\nonumber \\
&-& \frac{\lambda_{9}\lambda_{10}}{\lambda_{8}}-\frac{\lambda_{14}\lambda_{16}}{\lambda_{13}} \Big{)}\Big{(} \delta \rho^{*} \delta A_{1}^{'}+c.c. \Big{)}+ \Big{(} -\frac{b^4}{4\lambda_{8}}\e^2\rho^{'}\rho^{2}A_{x} -\frac{b^2}{2}\frac{\lambda_{6}\lambda_{15}}{\lambda_{8}\lambda_{13}}  \rho^{'}\Big{)}\Big{(}\delta \rho^{'} \delta A_{1}^{*} + c.c. \Big{)}\nonumber \\
&+& \Big{(} \frac{b^2}{2}\frac{\lambda_{10}}{\lambda_8} \rho^{'}- \frac{b^2}{2}\frac{\lambda_6 \lambda_{16}}{\lambda_8 \lambda_{13}} \rho^{'}\Big{)}
\Big{(}\delta \rho^{'*} \delta A_{1}^{ '} + c.c. \Big{)}
\ea
\ba
\label{L_{gamma-M}}
L_{\gamma M }& =& \Big{(}\frac{ab}{2}\e^2\rho^{2} A_{x}k_{x}k_{y}^2 + \frac{b\lambda_{12}}{2a\lambda_{13}}f^2 A_{x}^{'}k_{x}k_{y}^2 \Big{)} \Big{(}i\gamma  M^{*} + c.c. \Big{)} - \Big{(} \frac{b}{2a}A_{x}^{'}f^{2}k_{x}k_{y}^2\Big{)}\Big{(}i\gamma  M^{'*} + c.c. \Big{)}\nonumber\\
&+&\frac{1}{2\lambda_{2}}f^2k_{y}^2\Big{(}\lambda_{4}\gamma  M^{'*} + c.c. \Big{)}-\frac{\lambda_{7}}{2\lambda_{2}\lambda_{8}}f^2k_{y}^2\Big{(}\lambda_{3}\gamma  M^{'*} + c.c. \Big{)}-\frac{\lambda_{12}}{\lambda_{13}}\Big{(}\lambda_{17}^{*} \gamma  M^{'*} + c.c. \Big{)}\nonumber\\
&+&\Big{(}\frac{b\lambda_{11}}{2a\lambda_{13}}f^2 A_{x}^{'}k_{x}k_{y}^2\Big{)}\Big{(}i\gamma^{'}  M^{*} + c.c. \Big{)}-\Big{(}\frac{ab^3}{16\lambda_{1}\lambda_{2}}\e^2\rho^{2}f^2 A_{x}k_{x}^{3}k_{y}^4 \Big{)}\Big{(}i\gamma^{'}  M^{'*} + c.c. \Big{)} \nonumber\\ &-&\Big{(}\frac{ab\bar\lambda_{3}}{8\lambda_{2}\bar\lambda_{2}\lambda_{8}}f^2 k_{x}^{2}k_{y}^4 \Big{)}\Big{(}\lambda_{3}\gamma^{'}  M^{'*} + c.c. \Big{)}
-\Big{(}\frac{\lambda_{11}}{\lambda_{13}} \Big{)}\Big{(}\lambda_{17}^{*}\gamma^{'}  M^{'*} + c.c. \Big{)}
\ea
\ba
\label{L_{gamma-A1}}
L_{\gamma A_{1} }& =& \Big{(}\frac{b^2\lambda_{7}}{2\lambda_{8}}\e^2\rho^{2} A_{x}-\frac{\lambda_{12}\lambda_{15}}{\lambda_{13}}\Big{)} \Big{(}\gamma  \delta A_{1}^{*} + c.c. \Big{)} +\Big{(}\frac{b^2}{2a^2}\frac{i\lambda_{4}}{\lambda_{2}}f^{2}k_{x}-\frac{\lambda_{7}\lambda_{10}}{\lambda_{8}} -\frac{\lambda_{12}\lambda_{16}} {\lambda_{13}}\Big{)}\Big{(}\gamma \delta A_{1}^{'*} + c.c. \Big{)}\nonumber\\
&+&\Big{(}\frac{ab^3\bar\lambda_{3}}{8\bar\lambda_{2}\lambda_{8}}\e^2\rho^{2} A_{x}k_{x}^2k_{y}^2-\frac{\lambda_{11}\lambda_{15}}{\lambda_{13}}\Big{)}
\Big{(}\gamma^{'} \delta A_{1}^{*} + c.c. \Big{)}+\Big{(}\frac{b^5}{16a\lambda_{1}\lambda_{2}}\e^2\rho^{2} A_{x}f^2k_{x}^4k_{y}^2-\frac{ab\bar\lambda_{3}\lambda_{10}}{4\bar\lambda_{2}\lambda_{8}}k_{x}^2k_{y}^2\nonumber\\
&-&\frac{\lambda_{11}\lambda_{16}}{\lambda_{13}} \Big{)}\Big{(}\gamma^{'} \delta A_{1}^{'*} + c.c. \Big{)}
\ea
and
\ba
\label{L_{M-A1}}
L_{M A1}& =& \Big{(}-\frac{1}{2}f^2k_{x}k_{y}^2 + \frac{b\lambda_{15}}{2a\lambda_{13}}f^2 A_{x}^{'}k_{x}k_{y}^2 \Big{)} \Big{(}i\delta A_{1} M^{*} + c.c. \Big{)} - \Big{(} \frac{b^2}{4a^2\lambda_{2}}f^{4}k_{x}k_{y}^2\Big{)}\Big{(}i\delta A_{1}^{'*} M^{'} + c.c. \Big{)}\nonumber\\
&-&\frac{\lambda_{10}}{2\lambda_{2}\lambda_{8}}f^2k_{y}^2\Big{(}\lambda_{3}\delta A_{1}^{'} M^{'*} + c.c. \Big{)}-\Big{(}\frac{\lambda_{16}}{\lambda_{13}} \Big{)}\Big{(}\lambda_{17}^{*}\delta A_{1}^{'}  M^{'*} + c.c. \Big{)}
+\Big{(}\frac{b^2}{4\lambda_{2}\lambda_{8}}\e^2\rho^{2}A_{x}f^2k_{y}^2\Big{)}\nonumber\\
&&\Big{(}\lambda_{3}\delta A_{1} M^{'*} + c.c. \Big{)}-\frac{\lambda_{15}}{\lambda_{13}}\Big{(}\lambda_{17}^{*}\delta A_{1} M^{'*} + c.c. \Big{)}
+\frac{b\lambda_{16}}{2a\lambda_{13}}f^2A_{x}^{'}k_{x}k_{y}^2\Big{(}i\delta A_{1}^{'} M^{*} + c.c. \Big{)}
\ea
The parameters $\lambda_i$ and $\bar \lambda_i$ which are introduced after integrating out the non-dynamical fields  are defined via

\ba
\label{lambdak1}
\bar \lambda_1 &&= \frac{b^2}{2 a^2} k^2 f^2 + \frac{\e^2 }{2} b^2 \rho^2 \\
\bar \lambda_2 &&= \frac{a^2}{4} k_x^2 k_y^2 + \frac{\e^2 b^4 k_x^2 k^2}{4 a^2 \bar \lambda_1} f^2  \rho^2 A_x^2  \\
\bar \lambda_3 &&=  b b' k_x^2 - \frac{\e^2 b^4  k_x^2}{4 a^2 \bar \lambda_1}f^2 \rho^2 A_x A_x' \\
\bar \lambda_4 &&=  \frac{a b}{4} \left(\frac{a'}{a}-\frac{b'}{b} \right) k_x^2 k_y^2
+ \frac{\e^2 b^3 k_x^2 k_y^2}{4 a \bar \lambda_1} f^2 \rho^2 A_x A_x' \\
\bar \lambda_5 &&=- \frac{b^2}{2} \rho' k_x^2 + \frac{\e^2 b^4 k_x^2 }{2 \bar \lambda_1 a^2}
f f_{,\rho} \rho^2 A_x A_x'
\\
\lambda_1 &&=   \frac{a^2}{4} k_x^2 k_y^2 + \frac{\e^2 b^2 }{2}k_x^2 \rho^2 A_x^2 \\
\lambda_2 &&=   \frac{b^2 f^2}{2 a^2} k^2 + \frac{\e^2 b^2 a^2}{8 \lambda_1} \rho^2 k_x^2 k_y^2\\
\lambda_3 &&= - \frac{i k_x b^2}{2 a^2} f^2 A_x' + \frac{i \, \e^2  b^3 b'}{2\lambda_1}k_x^3 \rho^2 A_x \\
\lambda_4 &&= -\frac{a \lambda_3}{b} k_y^2 + \frac{i\, \e^2 a b^3}{8 \lambda_1}
  \left(\frac{a'}{a}+3\frac{b'}{b} \right)  k_x^3 k_y^2 \rho^2 A_x \\
 \lambda_5 &&=  \frac{i b^2}{a^2} k_x f f_{,\rho} A_x' - \frac{i\, \e^2 b^4}{4 \lambda_1}k_x^3\rho^2 \rho' A_x \\
 \lambda_6 && = \frac{a b}{2}  \left(\frac{a'}{a}+\frac{b'}{b} \right) k_y^2 + \frac{ a b\bar \lambda_3}{4 \bar \lambda_2} k_x^2 k_y^2
 \\
\lambda_7 && =  \frac{a b}{2} k_x^2 k_y^2 - \frac{a b^2 b'}{4 \lambda_1} \left(\frac{a'}{a}-\frac{b'}{b} \right) k_x^4 k_y^2 - \frac{\lambda_3^* \lambda_4}{\lambda_2}
\ea
\ba
  \lambda_8 && = - a^2 b^2 V - \frac{b^2 b'^2}{\lambda_1} k_x^4 - \frac{|\lambda_3|^2}{\lambda_2} - \frac{\e^2 b^2}{2} \rho^2 A_x^2
 \\
\lambda_9 && =  -\frac{a^2 b^2}{2} V_{,\rho} - \frac{b^2}{2 a^2} f f_{,\rho} A_x'^2 +
  \frac{b^3 b'}{2 \lambda_1}\rho' k_x^4 - \frac{\lambda_3^* \lambda_5}{\lambda_2}
  - \frac{\e^2 b^2}{2} \rho A_x^2
\\
\lambda_{10} && = - \frac{b^2}{2 a^2} f^2 A_x' - \frac{i  \, b^2 \lambda_3^* }{2 a^2 \lambda_2} f^2 k_x \\
\lambda_{11} && = -\frac{b^2}{4} k_x^2 k_y^2 - \frac{a^2 b^2}{16 \bar \lambda_2} k_x^4 k_y^4 - \frac{a b \bar \lambda_3 \lambda_6^*}{ 4 \bar \lambda_2\lambda_8} k_x^2 k_y^2
\\
\lambda_{12} && =  -\frac{b^2}{4} \left(\frac{a'}{a}-\frac{b'}{b} \right) k_x^2 k_y^2
+ \frac{a b \bar \lambda_4}{4 \bar \lambda_2} k_x^2 k_y^2 - \frac{\lambda_7 \lambda_6^*}{\lambda_8}
\\
\lambda_{13} && = \frac{b^2}{4} k_x^2 k_y^2 - \frac{a^2 b^2}{16 \bar \lambda_2} k_x^4 k_y^4 - \frac{| \lambda_6  |^2}{\lambda_8} \\
\lambda_{14} && = -\frac{a b}{2} \rho' k_y^2 + \frac{a b \bar \lambda_5}{4 \bar \lambda_2}
k_x^2 k_y^2 - \frac{\lambda_9 \lambda_6^*}{\lambda_8} \\
\lambda_{15} && =  - \frac{b}{2 a} f^2 A_x' k_y^2 + \frac{\e^2 b^2 \lambda_6^*}{2 \lambda_8} \rho^2 A_x \\
\lambda_{16} && =  -\frac{\lambda_6^* \lambda_{10}}{\lambda_8}
+ \frac{\e^2  b^5 }{16 a \bar \lambda_1 \bar \lambda_2 }f^2  \rho^2 A_x  k_x^4 k_y^2 \\
\label{lambda17}
\lambda_{17} && =  -\frac{\lambda_3^* \lambda_6^*}{2\lambda_2 \lambda_8} f^2 k_y^2
+\frac{i\, \e^2 a b^3}{16 \bar \lambda_1 \bar \lambda_2} f^2 A_x \rho^2 k_x^3 k_y^4
\ea
One can see that $\bar \lambda_i$ are determined by $\bar \lambda_1$ while $\lambda_i$
depends on both $\bar \lambda_1$ and $\lambda_1$. One can check from the detail processes of integrating out the non-dynamical fields that $\bar \lambda_1$ is obtained from integrating out $\delta A_0$. On the other hand, as we have seen 
in Section \ref{matter-leading}, the leading interactions originate from integrating out the matter sector. As a result, it is expected that $\bar \lambda_1$ plays the dominant role in determining the earliest time in which the interaction $\e^2 \rho^2 A_\mu A^\mu$ becomes comparable to $f^2 F^2$ interaction.  

Here we justify this conclusion specifically. To see this, let us look at $\lambda_1$. Dividing the second term in $\lambda_1$ to the first term in $\lambda_1$ yields 
\ba
\label{lambda1-ratio}
\frac{\e^2 \rho^2}{M_P^2 k^2} A_x^2  \, .
\ea
On the other hand, during most of period of inflation $\partial_t (\dot A f^2 e^{\alpha})=0 $
so $\dot A_x \sim e^{-\alpha } f^{-2} \sim e^{3\alpha}$. As a result $A_x \sim \dot A_x/3 H$.
Using the relation $\dot A_x^2 \sim (I \epsilon_H) e^{2 \alpha } f^{-2}$ from the attractor solution  we obtain  $A_x^2 \sim (I \epsilon_H) b^2 M_P^2/f^2$. Plugging this value of $A_x^2$ in the ratio Eq. (\ref{lambda1-ratio}) above yields 
\ba
(I \epsilon_H) \frac{\e^2 b^2 \rho^2}{k^2 f^2} \, .
\ea
Up to the pre-factor $I \epsilon_H \ll 1$ this ratio is the same as the ratio one obtains in comparing the second term in $\bar \lambda_1$ to the first term in $\bar \lambda_1$.
Now if we define ${\eta_c'}$ as the time when the second term in $\lambda_1$ becomes comparable  to the first term in $\lambda_1$, then $\eta_c \simeq (I \epsilon_H)^{-1/6}  \eta_c'$. Noting that $\eta<0$, we conclude that  $\eta_c \ll \eta_c' $. As a result the interaction $\e^2 \rho^2 A_\mu A^\mu$ becomes comparable to $f^2 F^2$ sooner in $\bar \lambda_1$ than in $\lambda_1$. Now, since the rest of $\lambda_i$ and $\bar \lambda_i$ are controlled by  either $\bar \lambda_1$ or $\lambda_1$, then we conclude that the earliest time when $\e^2 \rho^2 A_\mu A^\mu$  becomes comparable to $f^2 F^2$ is determined by $\bar \lambda_1$ as we used to fix $\eta_c$. \\ 

%%%%%%%%%%%%%%%%%%%%%%%%%%%%%%%%%%%%%%%%%%%%%%%%%%
\section{Second Order Slow-roll Action}
\label{second slow}

In this appendix we calculate the second order action for the canonical fields in the slow roll approximation. Following the discussions  in Section \ref{sec-ac} we divide the dynamical action into two different regions depending on whether the charge $\e$ is important or not.
In the first phase, $\eta < \eta_c$, the charge effect is sub-dominant while in the
second phase, $\eta > \eta_c$, its effect is dominant and the longitudinal mode, as we will introduce it in Eq. (\ref{D12}), has the same contribution as the transverse mode. In the following, first we write the action in the first phase and then we go to the second phase.

%%%%%%%%%%%%%%%%%%%%%%%%%%%%%%%%%%%%%%%%%%%%%%%%%
\subsubsection{First Phase}

Our goal is to write down the action in terms of the free Lagrangians plus the interaction terms.
Using the slow-roll approximation given in Eq. (\ref{slow roll}),  and taking $I \ll 1$ as mentioned in the main text, to leading orders in slow parameters and $I$ we have
\begin{align}
\label{Total scalar action1}
S_{2}^{(1)} = \int d\eta d^3k \bigg{(} L_{\rho \rho} + L_{\gamma\gamma} + L_{MM} + L_{A_{1}A_{1}} + L_{\rho\gamma} +   L_{\rho M} +  L_{\rho A_{1}} +  L_{\gamma A_{1}} +  L_{\gamma M} + L_{ A_{1} M} \bigg{)}~,
\end{align}
Where
\begin{align}
\label{Lrhorho}
L_{\rho \rho} & = \left(\frac{b^2}{2} \right) \Big| \delta \rho ^{'}\Big|^2 + \left(\frac{b^2}{2} \right)\bigg{(}-k^2+ (-\eta)^{-2}\bigg{(}6\epsilon_{H}-6\frac{\eta_{H}}{1-I}-12\frac{I}{1-I}(1-2\sin^2{\theta})\nonumber\\
&~~~+3\frac{\e^2 \cM^4}{k^2\lambda^2}\frac{a^2}{f^{2}}I\epsilon_{H}\cos^2{\theta}\bigg{)}\bigg{)}\Big| \delta \rho \Big|^2~~,\\
\label{Lgammagamma}
L_{\gamma \gamma} & = \left(\frac{b^2}{2a}k^2 \sin^2{\theta}\cos^2{\theta}\right)^2\bigg{(} 1 + 6\epsilon_{H}\bigg{)} \Big| \gamma ^{'}\Big|^2 -\left(\frac{b^2}{2a}k^3 \sin^2{\theta}\cos^2{\theta}\right)^2 \Big| \gamma \Big|^2~~,\\
\label{Lrhogamma}
L_{\rho \gamma} & = \left( \frac{\e^2}{8}\frac{b^3 \eta_{e}^4}{\eta^5}\frac{\cM^2}{\lambda^2 M_{P}}k^2 \cos^2{\theta} \sin^4{\theta} \sqrt{3\lambda} I \epsilon_{H}^{3/2} \right) \left( \delta \rho^{*} \gamma + c.c. \right) - \frac{3\sqrt{2}}{2} \frac{b^3}{a\eta}k^2 \cos^2{\theta} \sin^4{\theta} I\sqrt{\epsilon_{H}}\left( \delta \rho^{*} \gamma' + c.c. \right) \nonumber\\
&~~~-\frac{\sqrt{2}}{4} \frac{b^3}{a} k^2 \cos^2{\theta} \sin^2{\theta} \sqrt{\epsilon_{H}} \left( \delta \rho^{'*} \gamma' + c.c. \right) 
-\left( \frac{\e^2}{8}\frac{b^3 \eta_{e}^4}{\eta^6}\frac{\cM^2}{M_{P}\lambda^2} \cos^2{\theta} \sin^4{\theta} \sqrt{3\lambda}\left( 1 + 3\sqrt{2} \right) I \epsilon_{H}^{3/2} \right) \nonumber\\
&~~~\left( \delta \rho^{*} \gamma' + c.c. \right)
\end{align}
\begin{align}
\label{LMM}
L_{MM} & = \left(\frac{b^2}{2a^2}k^2f^2 \sin^2{\theta} \cos^2{\theta} + \frac{\e^2 \cM^4}{16\lambda^2}b^2 \sin^4{\theta}\frac{\epsilon_{H}}{1-I}\right)\Big| \cM^{'}\Big|^2 +\bigg{(}-\frac{b^2}{2a^2}k^4f^2 \sin^2{\theta} \cos^2{\theta}\nonumber\\
&~~~-\frac{\e^2 \cM^4}{16\lambda^2}b^2 k^2 \sin^2{\theta}\frac{\epsilon_{H}}{1-I}\bigg{)}\bigg{)}\Big| M \Big|^2~~,
\\
\label{LA1A1}
L_{A1 A1} & = \left(\frac{b^2}{2a^2}f^2 \sin^2{\theta} + \frac{\e^2 \cM^4}{16k^2\lambda^2}b^2 \cos^2{\theta}\frac{\epsilon_{H}}{1-I}\right)\Big| \delta A_{1}^{'}\Big|^2 +\bigg{(}-\frac{b^2}{2a^2}k^2f^2 \sin^2{\theta}\nonumber\\
&~~~-\frac{\e^2 \cM^4}{16\lambda^2}b^2\frac{\epsilon_{H}}{1-I}\bigg{)}\bigg{)}\Big|\delta A_{1} \Big|^2~~,
\\
\label{LrhoA1}
L_{\rho A1} & = \left(\frac{1}{\eta}\right) \left(\frac{fb^2}{a}\sqrt{6I} \sin^2{\theta} + \frac{\e^2 \cM^4}{8k^2\lambda^2}\frac {ab^2}{f} \epsilon_{H}\cos^2{\theta}\right)\Big{(} \delta \rho^{*}\delta A_{1}^{'} + c.c.\Big{)}-\e^2b^2 \rho A_{x}\Big{(}\delta \rho \delta A_{1}^{*} + c.c. \Big{)}~~,\\
\label{L_{rho-M}}
L_{\rho M} & = \left(\frac{1}{\eta}\right) \left(-\frac{fb^2}{a}\sqrt{6I} \sin^2{\theta}(k\cos{\theta}) + \frac{\e^2 \cM^4}{8k\lambda^2}\frac {ab^2}{f} \epsilon_{H} \sin^2{\theta}\cos{\theta}\right)\Big{(} \delta \rho^{*} M^{'} + c.c.\Big{)}~~\\
%\end{align}
%\begin{align}
\label{LMgamma}
L_{\gamma M} & = -\left(\frac{fb^3}{4a^2}k^5\sqrt{3I\epsilon_{H}}\sin^4{\theta}\cos^3{\theta}\right)\Big{(}i\gamma^{*}M + c.c.)\Big{)}-\left(\frac{b^3}{2a^2}\frac{f}{\eta}k^3\sqrt{3I\epsilon_{H}}\sin^2{\theta}\cos^3{\theta}\right)\nonumber\\
&\Big{(}i\gamma^{*}M^{'} + c.c.)\Big{)}
-\left(\frac{b^3}{4a^2}\frac{f}{\eta}k^3\sqrt{3I\epsilon_{H}}\sin^2{\theta}\cos^3{\theta}(1+\cos^2{\theta})\right)\Big{(}i\gamma^{'*}M + c.c.)\Big{)}\nonumber\\
&+\left(\frac{fb^3}{4a^2}k^3\sqrt{3I\epsilon_{H}}\sin^4{\theta}\cos^3{\theta}\right)\Big{(}i\gamma^{'*}M^{'} + c.c.)\Big{)}~~,\\
\label{LA1gamma}
L_{\gamma A_{1}} & =  \left(\frac{fb^3}{4a^2}k^4\sqrt{3I\epsilon_{H}}\sin^4{\theta}\cos^2{\theta}\right)\Big{(}\gamma^{*}\delta A_{1} + c.c.)\Big{)}+\left(\frac{b^3}{2a^2}\frac{f}{\eta}k^2\sqrt{3I\epsilon_{H}}\sin^2{\theta}\cos^2{\theta}\right)\nonumber\\
&\Big{(}\gamma^{*}\delta A_{1}^{'} + c.c.)\Big{)}
+\left(\frac{b^3}{4a^2}\frac{f}{\eta}k^2\sqrt{3I\epsilon_{H}}\sin^2{\theta}\cos^2{\theta}(1+\cos^2{\theta})\right)\Big{(}\gamma^{'*}\delta A_{1} + c.c.)\Big{)}\nonumber\\
&-\left(\frac{fb^3}{4a^2}k^2\sqrt{3I\epsilon_{H}}\sin^4{\theta}\cos^2{\theta}\right)\Big{(}\gamma^{'*}\delta A_{1}^{'} + c.c.)\Big{)}~~,\\
\label{LA1M}
L_{A_{1}M} & = \left(\frac{f^2b^2}{2a^2}k^3\sin^2{\theta}\cos{\theta}\right)\Big{(}i\delta A_{1}^{*}M + c.c.)\Big{)}+
\Big{(}-\frac{f^2b^2}{2a^2}k\sin^2{\theta}\cos{\theta}\nonumber\\
&+b^2\frac{\e^2 \cM^4}{16k\lambda^2}\frac{\epsilon_{H}}{1-I}\sin^2{\theta}\cos{\theta}\Big{)}\Big{(}iM^{'}\delta A_{1}^{'*} + c.c.)\Big{)}~~,
\end{align}
Now looking at  Eq. (\ref{LA1M}) we can easily see that this term is not as small as the other interaction terms. Actually this term seems to be of the same order as our free field action. This means that  $M $ and $\delta A_{1}$ are not the physical fields. One should consider a rotation in $ \{M, \delta A_{1} \}$ space such that all of the interaction terms become small compared to the free field action.
One can easily check that the following two new fields $D_1$ and $D_2$ work for us in the sense that they do not mix with each other and all of the interaction terms would be small:
\ba
\label{D12}
D_{1}&\equiv \delta A_{1} -ik \cos{\theta}M \\
D_{2}&\equiv \cos{\theta} \delta A_{1} + ik \sin^2{\theta}M \, .
\ea

One can check that  $D_{1}$ represents  the transverse polarization while $D_{2}$ is for the longitudinal polarization of the gauge field perturbations $\delta A_\mu$. Now the different parts of the action can be rewritten in terms of these two new fields  as
\begin{align}
L_{MM} +L_{\delta A_{1}\delta A_{1}} +L_{M\delta A_{1}} &= \frac{b^2 f^2}{2 a^2}\sin^2{\theta}\left( |D_{1}'|^2 -\left( k^2 + \frac{\e^2 \cM^4}{8 \lambda^2 } \frac{a^2}{f^2}\frac{\epsilon_{H}}{1-I} \right)|D_{1}|^2 \right)\nonumber\\
&+ \frac{\e^2 \cM^4}{16 \lambda^2 k^2}\frac{\epsilon_{H}b^2}{1-I}\left( |D_{2}'|^2 - k^2 |D_{2}|^2 \right)~~,\\
%\end{align}
%\begin{align}
\label{intD1-rho}
L_{\delta \rho M}+L_{\delta \rho \delta A_{1}} &= \left(\frac{1}{\eta}\right)\frac{b^2}{a} \sqrt{6I} \sin^2{\theta} f \Big{(} \delta \rho^{*} D_{1}^{'} + c.c.\Big{)} - \left( \frac{a^2}{f\eta} \right) \e^2 \sqrt{\frac{I \epsilon_{H}^2}{\lambda}}M_{P} \sin^2{\theta}\Big{(} \delta \rho^{*} D_{1} + c.c.\Big{)}\nonumber\\
& - \left( \frac{a^2}{f\eta} \right) \e^2 \sqrt{\frac{I \epsilon_{H}^2}{\lambda}}M_{P} \cos{\theta}\Big{(} \delta \rho^{*} D_{2} + c.c.\Big{)}~~,
%\end{align}
%\begin{align}
\end{align}
\begin{align}
\label{intD1-gamma}
L_{\gamma M}+L_{\gamma \delta A_{1}} &=
\left(\frac{b^3}{4a^2}k^4 f\sqrt{3I\epsilon_{H}}\sin^4{\theta}\cos^2{\theta}\right)\Big{(}\gamma^{*} D_{1} + c.c.)\Big{)}+\left(\frac{b^3}{2a^2}\frac{1}{\eta}k^2f\sqrt{3I\epsilon_{H}}\sin^2{\theta}\cos^2{\theta}\right)\nonumber\\
&\Big{(}\gamma^{*} D_{1}^{'} + c.c.)\Big{)}
+\left(\frac{b^3}{4a^2}\frac{1}{\eta}k^2f\sqrt{3I\epsilon_{H}}\sin^2{\theta}\cos^2{\theta}(1+\cos^2{\theta})\right)\Big{(}\gamma^{'*}D_{1} + c.c.)\Big{)}\nonumber\\
&-\left(\frac{b^3}{4a^2}k^2f\sqrt{3I\epsilon_{H}}\sin^4{\theta}\cos^2{\theta}\right)\Big{(}\gamma^{'*}\delta A_{1}^{'} + c.c.)\Big{)}~~ \, ,
\end{align}
where in Eq. (\ref{intD1-rho}) and Eq. (\ref{intD1-gamma}) only the leading terms have been written. It can easily be seen that since the charge effect is not important in this phase, the longitudinal mode is sub-leading.
Now we can define the canonical variable as,
\begin{align}
\label{canonical variables}
\overline{\delta \rho} &\equiv b \delta \rho \\
\overline{\gamma}&\equiv \frac{a}{\sqrt{2}}\frac{k_{x}^2 k_{y}^2}{k^2} \gamma \\
\overline{D_{1}}&\equiv \frac{b}{a}f \sin{\theta} D_{1} \\
\overline{D_{2}}&\equiv \frac{\e \cM^2}{2\sqrt{2}\lambda M_{P} k}\sqrt{\frac{\epsilon_{H}}{1-I}} bD_{2}
\end{align}
One can write the action in terms of the canonical variables. Due to technical reasons we only express the free Lagrangians in terms of the canonical variables while the
interaction terms may be expressed in terms of the old variables
\begin{align}
\label{rho rho action}
L_{\rho \rho} &= \frac{1}{2}|\overline{\delta \rho}'|^2 + \frac{1}{2}\bigg{[} -k^2 + (-\eta)^{-2}\bigg{(}2+9\epsilon_{H}-6\frac{\eta_{H}}{1-I}
-12\frac{I}{1-I}(1-2\sin^2{\theta})\bigg{)}\bigg{]}|\overline{\delta \rho}|^2\\
\label{gamma gamma action}
L_{\gamma \gamma} &= \frac{1}{2}|\overline{\gamma}^{'}|^2 + \frac{1}{2}\bigg{[} -k^2 + (-\eta)^{-2}\bigg{(}2 + 15\epsilon_{H} +I\epsilon_{H}(-6+11\sin^2{\theta} + \cot^2{\theta})\bigg{)}\bigg{]}|\overline{\gamma}|^2 \\
\label{D1 D1  action}
L_{D_{1} D_{1}} &= \frac{1}{2}|\overline{D_{1}}^{'}|^2 + \frac{1}{2}\bigg{[} -k^2 + (-\eta)^{-2}\bigg{(}2+9\epsilon_{H}-3\frac{\eta_{H}}{1-I}\bigg{)}\bigg{]}|\overline{D_{1}}|^2\\
\label{D2 D2  action}
L_{D_{2} D_{2}} &= \frac{1}{2}|\overline{D_{2}}^{'}|^2 + \frac{1}{2}\bigg{[} -k^2 + (-\eta)^{-2}\bigg{(}2+3\epsilon_{H} + I\epsilon_{H}\bigg{)}\bigg{]}|\overline{D_{2}}|^2\\
\end{align}
These are the final results for the action in the first phase used in Eqs. (\ref{rhorho action1})-
(\ref{D2D2 action1}).

%%%%%%%%%%%%%%%%%%%%%%%%%%%%%%%%%%%%%%%%%%%%%%%%%%
\subsubsection{Second Phase}
In this part we  write the leading order action in the second phase from which we can read off the canonical fields. Considering the dominant effects of $\e$ in $\bar \lambda_1$ as mentioned in Section {\ref{sec-ac}} yields
\begin{align}
\label{rho rho action2}
L_{\rho \rho} &= \frac{b^2}{2}|\delta \rho'|^2 -\left(\frac{\e^2 I \epsilon_{H} \lambda }{\cM^4}\right) \left(\frac{b^2}{f^{2}\eta^{2}}\right)|\delta \rho|^2\\
\label{D1 D1 action2}
L_{D_{1} D_{1}} &= \frac{b^2}{2a^2}f^2\sin^2{\theta}|D_{1}'|^2 -\left(\frac{ a^2 \e^2 \epsilon_{H} \cM^4}{16\lambda^2 f^2}\right)\left(\frac{b^2}{a^2}f^2\sin^2{\theta}\right)|D_{1}|^2\\
\label{D2 D2 action2}
L_{D_{2} D_{2}} &= \frac{b^2}{2a^2}f^2|D_{2}'|^2 -\left(\frac{ a^2 \e^2 \epsilon_{H}\cM^4}{16\lambda^2 f^2}\right)\left(\frac{b^2}{a^2}f^2\right)|D_{2}|^2\\
\label{rhoD1 action2}
L_{\rho D_{1}} &= \left(\frac{1}{\eta}\right)\frac{b^2}{a} \sqrt{6I} \sin^2{\theta} f \Big{(} \delta \rho^{*} D_{1}^{'} + c.c.\Big{)} - \left( \frac{a^2}{f\eta} \right) \e^2 \sqrt{\frac{I \epsilon_{H}^2}{\lambda}}M_{P} \sin^2{\theta}\Big{(} \delta \rho^{*} D_{1} + c.c.\Big{)}\\
\label{rhoD2 action2}
L_{\rho D_{2}} &= \left(\frac{1}{\eta}\right)\frac{b^2}{a} \sqrt{6I} \cos{\theta} f \Big{(} \delta \rho^{*} D_{2}^{'} + c.c.\Big{)}- \left( \frac{a^2}{f\eta} \right) \e^2 \sqrt{\frac{I \epsilon_{H}^2}{\lambda}}M_{P} \sin^2{\theta}\Big{(} \delta \rho^{*} D_{2} + c.c.\Big{)}\\
L_{\rho \gamma} & = \left( \frac{\e^2}{8}\frac{b^3 \eta_{e}^4}{\eta^5}\frac{\cM^2}{\lambda^2 M_{P}}k^2 \cos^2{\theta} \sin^4{\theta} \sqrt{3\lambda} I \epsilon_{H}^{3/2} \right) \left( \delta \rho^{*} \gamma + c.c. \right) - \frac{3\sqrt{2}}{2} \frac{b^3}{a\eta}k^2 \cos^2{\theta} \sin^4{\theta} I\sqrt{\epsilon_{H}}\times  \nonumber\\
&~~~ \times \left( \delta \rho^{*} \gamma' + c.c. \right)-\frac{\sqrt{2}}{4} \frac{b^3}{a} k^2 \cos^2{\theta} \sin^2{\theta} \sqrt{\epsilon_{H}} \left( \delta \rho^{'*} \gamma' + c.c. \right) 
\end{align}
Now from the above equations, we can find the canonical variables in the second phase as,
\begin{align}
\label{canonical variables2 appen}
\overline{\delta \rho}_{k} &\equiv b \delta \rho_{k} \\
\overline{D_{1k}}&\equiv \frac{b}{a} \sin{\theta} f D_{1k} \\
\overline{D_{2k}}&\equiv \frac{b}{a} f D_{2k}
\end{align}
as used in Eq. (\ref{canonical variables2}).

%%%%%%%%%%%%%%%%%%%%%%%%%%%%%%%%%%%%%%%%%%%%%%%%%%
\section{Outgoing Mode Functions}
\label{outgoing modes}

In this appendix we calculate the mode function during the second phase, $\eta< \eta_c < \eta_e$.

The canonical inflaton field mode function, $\overline{\delta \rho}_{\textbf{k}}$, associated
with the free inflaton Lagrangian given in Eq. (\ref{rhorho action})  is
\ba
\label{mode function-u}
\overline{\delta \rho}_{\textbf{k}} &=& u^{(m)}_{\textbf{k}}a_{\rho\textbf{k}} + u^{(m)*}_{(-\textbf{k})}a^{\dag}_{\rho(-\textbf{k})} \nonumber \\
u^{m}_{\textbf{k}} &=&\sqrt{-\eta}\left[ c_{1k}H_{3/4}^{(1)}\left( \frac{\sqrt{\Omega}}{2\eta^2 }\right) + c_{2k}H_{3/4}^{(2)}\left( \frac{\sqrt{\Omega}}{2\eta^2 }\right)\right]~~~,~~~ \Omega \equiv \frac{2\e^2I\epsilon_{H}\lambda M_P^4}{\cM^4}\eta_{e}^4.
\ea
Here $H_{3/4}^{(1, 2)} ( x)$ are the Hankel functions with index $\frac{3}{4}$ and $c_{i \,  k}$ are constant of integrations to be found by matching conditions at $\eta =\eta_c$.

Similarly the canonical transverse mode function,  $\overline{D_{1\textbf{k}}}$, associated with the Lagrangian $L_{D_1 D_1}$ in Eq. (\ref{D1D1 action}) is
\ba
\label{mode function-v}
\overline{D_{1\textbf{k}}} &=& \left(\frac{b}{a} \sin{\theta} f\right) \left(v^{(m)}_{\textbf{k}}b_{\rho\textbf{k}} + v^{(m)*}_{(-\textbf{k})}b^{\dag}_{\rho(-\textbf{k})} \right) \nonumber \\
v^{m}_{\textbf{k}} &=&\left( \frac{a \sqrt{-\eta}}{b\sin{\theta} f}\right)\left[ d_{1k}H_{3/4}^{(1)}\left( \frac{\sqrt{\Delta}}{2\eta^2 }\right) + d_{2k}H_{3/4}^{(2)}\left(\frac{\sqrt{\Delta}}{2\eta^2 }\right)\right]~~~,~~~ \Delta \equiv \frac{3\e^2\epsilon_{H}}{2\lambda}\eta_{e}^4.
\ea
Finally the $\overline{D_{2\textbf{k}}}$ mode function is
\ba
\label{mode function-w}
\overline{D_{2\textbf{k}}} &=&\left(\frac{b}{a}f\right) \left( w^{(m)}_{\textbf{k}}c_{\rho\textbf{k}} + w^{(m)*}_{(-\textbf{k})}c^{\dag}_{\rho(-\textbf{k})} \right)\nonumber \\
w^{(m)}_{\textbf{k}} &=&\left(\frac{a \sqrt{-\eta}}{bf}\right) \left[ e_{1k}H_{3/4}^{(1)}\left( \frac{\sqrt{\Delta}}{2\eta^2 }\right) + e_{2k}H_{3/4}^{(2)}\left( \frac{\sqrt{\Delta}}{2\eta^2 }\right)\right].
\ea
Our goal is to find the constants of integration $c_{ik},d_{ik},e_{ik}~ ,~ i=1,2$ with the appropriate matching conditions at $\eta = \eta_c$.
To impose the matching conditions we require that the original fields $\delta \rho, M, A_1$ and their derivatives  to be continuous at $\eta= \eta_c$. With the incoming mode functions given by Eq. (\ref{mode function0}) and after  imposing the matching conditions one obtains
\begin{align}
%\label{Coefficients}
\label{c12}
c_{1,2} &= \frac{\pi}{8i}\frac{e^{-ik\eta_{c}}}{\sqrt{-2k\eta_{c}}}\left(\pm H^{(2,1)}_{3/4}\left(\frac{\sqrt{\Omega}}{2\eta_{c}^2 }\right)\left(3-\frac{3i}{k\eta_{c}}+ ik\eta_{c}\right)\mp H^{(2,1)}_{-1/4}\left(\frac{\sqrt{\Omega}}{2\eta_{c}^2 }\right)\frac{\sqrt{\Omega}}{\eta_{c}^2 }\left(1-\frac{i}{k\eta_{c}}\right)\right),\\
\label{d12}
d_{1,2}& = \frac{\pi}{8i}\frac{e^{-ik\eta_{c}}}{\sqrt{-2k\eta_{c}}}\left(\pm H^{(2,1)}_{3/4}\left(\frac{\sqrt{\Delta}}{2\eta_{c}^2 }\right)\left(3-\frac{3i}{k\eta_{c}}+ ik\eta_{c}\right)\mp H^{(2,1)}_{-1/4}\left(\frac{\sqrt{\Delta}}{2\eta_{c}^2 }\right)\frac{\sqrt{\Delta}}{\eta_{c}^2 }\left(1-\frac{i}{k\eta_{c}}\right)\right),\\
\label{e12}
e_{1,2} &= \frac{\pi}{8i}\frac{e^{-ik\eta_{c}}}{\sqrt{-2k\eta_{c}}}\left(\pm H^{(2,1)}_{3/4}\left(\frac{\sqrt{\Delta}}{2\eta_{c}^2 }\right)\left( ik\eta_{c}\right)\mp H^{(2,1)}_{-1/4}\left(\frac{\sqrt{\Delta}}{2\eta_{c}^2 }\right)\frac{\sqrt{\Delta}}{\eta_{c}^2 }\left(1-\frac{i}{k\eta_{c}}\right)\right).
\end{align}
This fixes the form of the outgoing mode functions. However, as discussed at the end of Section \ref{sec-ac}, during most of the period of the second phase the arguments of the Hankel functions above are smaller than unity so the mode functions to a good approximation follow the profile of a massless scalar field. As a result, in our In-In integrals we can use the mode functions as given in Eq. (\ref{mode function0}).

%%%%%%%%%%%%%%%%%%%%%%%%%%%%%%%%%%%%%%%%%%%%%%%%%%
%\newpage

\end{document}